\documentclass[reprint,superscriptaddress,nofootinbib,amsmath,amssymb,aps,onecolumn,notitlepage]{revtex4-1}

\usepackage[a4paper, margin=2.50cm]{geometry}
\usepackage{epsfig}
\usepackage{amsmath}
\usepackage{graphicx}
\usepackage{color}
\usepackage{amsfonts,amssymb}
\usepackage{array}
\usepackage{tabu}
\usepackage[utf8]{inputenc}
\usepackage[english]{babel}
\usepackage[T1]{fontenc}
\usepackage{lmodern}
\usepackage{microtype}
\usepackage{amsthm}
\usepackage{dsfont}
\usepackage[utf8]{inputenc}
\usepackage[english]{babel}
\usepackage[T1]{fontenc}
\usepackage{graphicx}
\usepackage{amsmath}
\usepackage{amsfonts}
\usepackage{amssymb}
\usepackage{makecell}
\usepackage{dsfont}
\usepackage{microtype}
\usepackage[breaklinks=true,colorlinks]{hyperref}
\usepackage[utf8]{inputenc}
\usepackage[english]{babel}
\usepackage[T1]{fontenc}
\usepackage{graphicx}
\usepackage{amsmath}
\usepackage{amsfonts}
\usepackage{amssymb}
\usepackage{dsfont}

\newcommand{\be}{\begin{equation}}
\newcommand{\ee}{\end{equation}}
\newcommand{\ba}{\begin{eqnarray}}
\newcommand{\ea}{\end{eqnarray}}

\newcommand{\cf}{cf.\ }
\newcommand{\coloneq}{\mathrel{\mathop:}=}
\newcommand{\eqcolon}{=\mathrel{\mathop:}}
\newcommand{\dd}{\mathrm{d}}
\newcommand{\Tr}{\operatorname{Tr}}

\newcommand{\kB}{k_\mathrm{B}}
\newcommand{\Treal}{T_\mathrm{h(real)}}
\newcommand{\nbar}{\bar{n}}

\newcommand{\Srel}[2]{S\left({#1}\middle\|{#2}\right)}
\newcommand{\bea}{\begin{eqnarray}}
\newcommand{\eea}{\end{eqnarray}}

\newcommand{\nn}{\nonumber}

\begin{document}

\title{Thermodynamic principles and implementations of quantum machines}
\author{Arnab Ghosh}
\affiliation{Department of Chemical and Biological Physics, Weizmann Institute of Science, Rehovot 7610001, Israel}

\author{Wolfgang Niedenzu}
\affiliation{Department of Chemical and Biological Physics, Weizmann Institute of Science, Rehovot 7610001, Israel}
\affiliation{Institut f\"ur Theoretische Physik, Universit\"at 
Innsbruck, Technikerstra{\ss}e 21a, A-6020~Innsbruck, Austria}

\author{Victor Mukherjee}
\affiliation{Department of Chemical and Biological Physics, Weizmann Institute of Science, Rehovot 7610001, Israel}

\author{Gershon Kurizki}
\email{gershon.kurizki@weizmann.ac.il} 
\affiliation{Department of Chemical and Biological Physics, Weizmann Institute of Science, Rehovot 7610001, Israel}

\date{November 11, 2018}

\begin{abstract}
The efficiency of cyclic heat engines is limited by the Carnot bound. This bound follows from the second law of thermodynamics and is attained by engines that operate between two thermal baths under the reversibility condition whereby the total entropy does not increase. By contrast, the efficiency of engines powered by quantum non-thermal baths has been claimed to surpass the thermodynamic Carnot bound. The key to understanding the performance of such engines is a proper division of the energy supplied by the bath to the system into heat and work, depending on the associated change in the system entropy and ergotropy. Due to their hybrid character, the efficiency bound for quantum engines powered by a non-thermal bath does not solely follow from the laws of thermodynamics. Hence, the thermodynamic Carnot bound is inapplicable to such hybrid engines. Yet, they do not violate the principles of thermodynamics.

\par

An alternative means of boosting machine performance is the concept of heat-to-work conversion catalysis by quantum non-linear (squeezed) pumping of the piston mode. This enhancement is due to the increased ability of the squeezed piston to store ergotropy. Since the catalyzed machine is fueled by thermal baths, it adheres to the Carnot bound.

\par 

We conclude by arguing that it is not quantumness per se that improves the machine performance, but rather the properties of the baths, the working fluid and the piston that boost the ergotropy and minimize the wasted heat in both the input and the output.

\end{abstract}
\maketitle

\tableofcontents

\newpage

\section{Introduction}

Engines transform some form of energy, such as thermal, chemical, mechanical, or electrical energy into useful work. Their efficiency, namely, the ratio of the output work to the input energy, is restricted to $1$ at most by energy conservation. Engines converting mechanical (potential, kinetic or rotational) energy into work may, in principle, approach unit efficiency. By contrast, the efficiency of heat-to-work conversion in a cyclic heat engine that operates between cold and hot thermal baths with temperature ratio $T_\mathrm{c}/T_\mathrm{h}$ is independent of the specific design and limited by the universal Carnot bound~\cite{carnotbook}. This bound follows from the second law of thermodynamics under the \textit{reversibility} condition, whereby the total (combined) entropy of the engine and the two baths is unaltered over a cycle~\cite{kondepudibook}.

\par

As opposed to standard heat engines (HE) conforming to the above description, diverse models of cyclic engines energized by \textit{quantum non-thermal} baths have been suggested to \textit{surpass} the Carnot bound~\cite{scully2003extracting,dillenschneider2009energetics,huang2012effects,abah2014efficiency,rossnagel2014nanoscale,hardal2015superradiant,niedenzu2016operation,manzano2016entropy,klaers2017squeezed,agarwalla2017quantum}. Until recently, there has been no clear, rigorously founded answer to the questions: Is there a common mechanism for such surpassing? And if there is, does it not violate the second law? The issue is even broader: In many models the engine comprises quantum-mechanical ingredients whose purpose is to provide a ``quantum advantage'' or ``quantum supremacy''~\cite{scully2003extracting,dillenschneider2009energetics,huang2012effects,abah2014efficiency,
rossnagel2014nanoscale,hardal2015superradiant,niedenzu2016operation,manzano2016entropy,klaers2017squeezed,agarwalla2017quantum,
alicki2004thermodynamics,quan2007quantum,gemmerbook,linden2010how, dorner2012emergent, abah2012single, gelbwaser2013minimal, kosloff2013quantum, delcampo2014more, gelbwaser2014heat, kosloff2014quantum, skrzypczyk2014work, binder2015quantum, gelbwaser2015thermodynamics, uzdin2015equivalence, uzdin2016quantum, vinjanampathy2016quantum,rossnagel2016single,klatzow2017experimental}. What are, if any, their common operational principles and their impact on the performance? In particular, can we assess their maximum efficiency from the second law via the reversibility condition as in standard HE?

\par

Here we present an overview of our recent endeavor~\cite{niedenzu2016operation,dag2016multiatom,niedenzu2018quantum,ghosh2017catalysis} to resolve the foregoing principle issues of cyclic quantum machines fuelled by arbitrary baths: The fundamental aspect of this endeavor has been the understanding of the role played by the first and second laws in quantum dissipative processes that characterize bath-powered engines. It is widely accepted~\cite{alicki1979quantum,kosloff1984quantum,alicki2004thermodynamics,boukobza2007three,parrondo2009entropy,
deffner2011nonequilibrium,boukobza2013breaking,kosloff2013quantum,sagawa2013second,argentieri2014violation,
binder2015quantum,gelbwaser2015thermodynamics,manzano2016entropy,
vinjanampathy2016quantum,brandner2016periodic,breuerbook} that the second law applied to quantum relaxation processes is faithfully rendered by Spohn's inequality~\cite{spohn1978entropy}. According to this inequality, the entropy change of a system that interacts with a thermal bath is bounded from below by the exchanged energy divided by the bath temperature. We have shown~\cite{niedenzu2018quantum}, however, that the bound on entropy change in quantum dissipative processes crucially depends on whether the state of the system is non-passive. The definition of a non-passive state~\cite{pusz1978passive,lenard1978thermodynamical,allahverdyan2004maximal} is that its energy can be unitarily reduced until the state becomes passive, thereby extracting work. The maximum amount of work extractable from such states (their ``work capacity'') has been dubbed ``ergotropy'' in Ref.~\cite{allahverdyan2004maximal}. Non-passive states may thus be thought of as ``quantum batteries''~\cite{alicki2013entanglement,binder2015quantacell,friis2018precisionandwork}. Examples of non-passive states are squeezed or coherent states of a harmonic oscillator~\cite{gelbwaser2014heat,gelbwaser2013work,niedenzu2016operation}. 

\par

The significance of non-passivity~\cite{pusz1978passive,lenard1978thermodynamical,allahverdyan2004maximal,
anders2013thermodynamics,alicki2013entanglement,gelbwaser2013work,
binder2015quantacell,binder2015quantum,gelbwaser2015thermodynamics,
skrzypczyk2015passivity,brown2016passivity,dag2016multiatom,depalma2016passive,
niedenzu2016operation,vinjanampathy2016quantum,friis2018precisionandwork} as a work resource in bath-powered engines has been demonstrated by us in our classification~\cite{niedenzu2016operation,niedenzu2018quantum,dag2016multiatom} of such machines according to their operation principle, as discussed below:

\par 

\begin{itemize}
\item \emph{Machines of the first kind} are fuelled by a non-thermal bath, such as a squeezed thermal~\cite{rossnagel2014nanoscale} or coherently-displaced thermal bath~\cite{brown2016passivity}, that render the working fluid (WF) steady-state \emph{non-passive}~\cite{pusz1978passive,lenard1978thermodynamical,allahverdyan2004maximal,binder2015quantum,
gelbwaser2013work,gelbwaser2014heat,gelbwaser2015thermodynamics}. We have pointed out~\cite{niedenzu2016operation,niedenzu2018quantum,dag2016multiatom} that in such machines the energy imparted by the non-thermal bath consists of a part that increases the ergotropy of the WF and an entropy-changing part that changes the WF's passive energy. This division of the transferred energy into work-like and heat-like contributions implies that the efficiency of machines of the first kind \textit{does not have a thermodynamic bound that may be deduced from the reversibility condition expressed by the second law.} This becomes clear when the energy of the non-thermal bath has no thermal component and is pure ergotropy: The engine can then deliver work without heat input. Hence, such machines may be thought of as \emph{hybrid thermo-mechanical engines}. Their efficiency bound cannot be properly compared with the Carnot bound, since the latter is a restriction imposed by the second law on heat~\cite{kondepudibook} but not on work imparted by the bath.

\par

\item \emph{Machines of the second kind} are heat engines wherein the WF is \textit{thermalized}, i.e., is rendered passive by the bath, be it thermal or non-thermal. Hence, the entire energy transfer from the bath to the WF corresponds to heat. In addition to standard HE, machines of the second kind are exemplified by the pioneering model introduced by Scully and co-workers~\cite{scully2003extracting} of a Carnot heat engine powered by partly-coherent three-level atoms (dubbed ``phaseonium'') that interact with a cavity-mode WF as they cross the cavity. The surprising finding~\cite{scully2003extracting} was that a beam of such ``phaseonium'' atoms may be treated as a non-thermal, quantum-coherent bath that, for an appropriate phase $\varphi$ of the interlevel coherence, can thermalize the cavity field to a temperature $T_\varphi>T$, where the latter is the atoms' temperature without coherence. Machines of the second kind act as a genuine heat engine, whose efficiency is limited by the Carnot bound, but the one corresponding to the \emph{real} temperature of the WF, i.e., $T_\varphi$. 
\end{itemize}

Instead of quantum machines fueled by non-thermal baths~\cite{niedenzu2016operation,dag2016multiatom}, which adhere to rules that differ from those of quantum heat engines, we may also consider machines fueled by thermal baths but still allow for an enhanced ergotropy of the piston state. We show that the non-linear pumping of the piston mode provides a powerful boost on the performance of the machine~\cite{ghosh2017catalysis}, i.e., its output power and efficiency which is determined by the capacity of the piston state to store work, i.e., its ergotropy. Finally, we discuss the question: Is the engine performance in the above scenarios determined by quantumness, i.e., the quantum features of the setup? The topics outlined above are to be discussed in detail in the sections that follow.  

\section{Energy exchange between a driven system and a bath}

The distinction discussed above between machines of first and second kind is due to the different nature of their system-energy exchange in the quantum domain. Consider an initially prepared state $\rho_0$ of a quantum system that evolves into a state $\rho(t)$ under the action of a Hamiltonian $H(t)$ and a bath. In general, the bath and/or the system may be in a non-thermal state. The change in the system energy $E(t)=\Tr[\rho(t)H(t)]$ is decomposed as follows~\cite{alicki1979quantum,kosloff1984quantum},
\begin{equation}\label{eq_first_law}
  \Delta E(t)=W(t)+\mathcal{E}_\mathrm{d}(t).
\end{equation}
Here the first term 
\begin{subequations}\label{eq_defs_Ediss_work}
  \begin{equation}\label{eq_def_work}
    W(t)\coloneq\int_0^t\Tr[\rho(t^\prime)\dot H(t^\prime)]\dd t^\prime
  \end{equation}
is the work~\cite{pusz1978passive} that is is either extracted or invested by an external source, as in driven engines~\cite{gelbwaser2015thermodynamics}. The second term 
\begin{equation}\label{eq_def_DeltaEdiss}
    \mathcal{E}_\mathrm{d}(t)\coloneq\int_0^t\Tr[\dot\rho(t^\prime)H(t^\prime)]\dd t^\prime
  \end{equation}
\end{subequations} 
is the energy change of the system due to its dissipative interaction with the bath. 

\par

We now further decompose $\mathcal{E}_\mathrm{d}(t)$ according to the associated entropy change, which naturally relates to the concept of non-passive states and ergotropy~\cite{pusz1978passive,lenard1978thermodynamical,allahverdyan2004maximal}: The energy $E$ of a state $\rho$ can be decomposed into \textit{ergotropy} $\mathcal{W}\geq0$ and \textit{passive energy} $E_\mathrm{pas}$. Ergotropy is defined as the maximum amount of work extractable from $\rho$ by means of unitary transformations and reads
\begin{equation}\label{eq_app_def_ergotropy}
  \mathcal{W}(\rho,H)\coloneq\Tr(\rho H)-\min_U\Tr(U\rho U^\dagger H)\geq0,
\end{equation}
where the minimization is over all possible unitary transformations $U$. Passive energy, by contrast, cannot be extracted as useful work in a cyclic, unitary, fashion. For a state $\rho$, there is a passive state $\pi=V \rho V^\dagger$ that only contains passive energy, where $V$ is the unitary that minimizes the second term on the r.h.s.\ of Eq.~\eqref{eq_app_def_ergotropy}. The corresponding passive energy is $E_\mathrm{pas}=\Tr[V \rho V^\dagger H]=\Tr[\pi H]$ such that the energy of the state $\rho$ can be written as
\begin{equation}
  E=E_\mathrm{pas}+\mathcal{W}=\Tr[\pi H]+\Tr[(\rho-\pi)H].
\end{equation}

\par

Since $\rho$ and $\pi$ are related by a unitary transformation, their entropies $\mathcal{S}(\rho)=\mathcal{S}(\pi)$ coincide. Here $\mathcal{S}(p)=-\kB\Tr[p\ln p]$ is the von~Neumann entropy of the state $p$. This observation motivates the decomposition of the dissipative energy change~\eqref{eq_def_DeltaEdiss} as    
\begin{equation}\label{eq_DeltaEdiss_decomposition}
  \mathcal{E}_\mathrm{d}(t)=\mathcal{Q}(t)+\Delta\mathcal{W}|_\mathrm{d}(t).
\end{equation}
Here
\begin{subequations}\label{eq_defs_heat_passive_work}
  \begin{equation}\label{eq_def_heat}
    \mathcal{Q}(t)\coloneq\int_{0}^{t} \Tr[\dot{\pi}(t^\prime)H(t^\prime)]\dd t^\prime
  \end{equation}
  is associated with a change in the passive state and thus a change in entropy. In analogy to thermodynamics, due to its entropy-changing character, we refer  to Eq.~\eqref{eq_def_heat} as \textit{heat}. While $\mathcal{Q}(t)$ is the dissipative (non-unitary) change in \textit{passive energy}, the second contribution in Eq.~\eqref{eq_DeltaEdiss_decomposition},
  \begin{equation}\label{eq_def_DeltaW_diss}
    \Delta\mathcal{W}|_\mathrm{d}(t)\coloneq\int_0^t\Tr\Big[\big(\dot\rho(t^\prime)-\dot\pi(t^\prime)\big)H(t^\prime)\Big]\dd t^\prime,
  \end{equation}
\end{subequations}
is the dissipative (non-unitary) change in the system \textit{ergotropy} due to its interaction with the bath. If the system state is always passive, Eqs.~\eqref{eq_def_heat} and~\eqref{eq_def_DeltaEdiss} coincide.

\par

Hence, the system ergotropy may increase in a non-unitary fashion due to interaction with a bath and be subsequently extracted from the system in the form of work via a suitable unitary process~\cite{alicki2017gkls}. Any \textit{unitary} change in passive energy due to the time-dependence of the Hamiltonian contributes to the work~\eqref{eq_def_work}. The decomposition of the exchanged energy~\eqref{eq_DeltaEdiss_decomposition} into dissipative changes in passive energy~\eqref{eq_def_heat} and ergotropy~\eqref{eq_def_DeltaW_diss}, is a new unraveling of the first law of thermodynamics for quantum systems~\cite{niedenzu2018quantum}. 

\par 

The energies~\eqref{eq_defs_heat_passive_work} are, in general, process variables and thus depend on the evolution path. By contrast, for a constant Hamiltonian, they become path-independent and reduce to the change in passive energy
\begin{subequations}
\begin{equation}\label{eq_Q_t} 
\mathcal{Q}(t)=\Delta E_\mathrm{pas}(t)=\Tr[\pi(t)H]-\Tr[\pi_0H],
\end{equation}
and the change in system ergotropy,
\begin{equation}\label{eq_W_t}
\Delta\mathcal{W}|_\mathrm{d}(t)=\Delta\mathcal{W}(t)=\mathcal{W}(\rho(t))-\mathcal{W}(\rho_0),
\end{equation}
\end{subequations}
respectively.

\section{Reversibility and Entropy production}\label{sec_spohn}

In this section we derive the bound on the entropy change of a quantum system that is in contact with a (thermal or non-thermal) bath~\cite{niedenzu2018quantum} based on our above analysis~\cite{uzdin2018second}. A tight estimate of entropy change is required to deduce the maximum work obtainable from cyclic quantum machines. We use the division~\eqref{eq_defs_heat_passive_work} discussed above to derive a bound that is much tighter than that obtained from the second law (non-negative entropy production)~\cite{kondepudibook} in scenarios where non-passive states of the WF are involved.  

In non-equilibrium thermodynamics, a dissipative processes is irreversible if the total entropy of the system and the bath combined increases (positivity of its entropy production)~\cite{kondepudibook}. This criterion translates into Spohn's inequality for the entropy production rate~\cite{spohn1978entropy} by quantum systems that are weakly coupled to thermal or non-thermal baths,
\begin{equation}\label{eq_spohn}
  \sigma\coloneq-\frac{\dd}{\dd t}\Srel{\rho(t)}{\rho_\mathrm{ss}}\geq0,
\end{equation}
where $\Srel{\rho(t)}{\rho_\mathrm{ss}}\coloneq\kB\Tr[\rho(t)(\ln \rho(t)-\ln \rho_\mathrm{ss})]$ is the system entropy relative to its steady state $\rho_\mathrm{ss}$. Inequality~\eqref{eq_spohn} holds for any $\rho(t)$ that evolves according to a Markovian master equation~\cite{breuerbook}
\begin{equation}\label{eq_app_master}
  \dot\rho=\mathcal{L}\rho,
\end{equation}
where $\mathcal{L}$ is a Lindblad operator, the steady-state solution obeying $\mathcal{L}\rho_\mathrm{ss}=0$. The time-integrated ($t \rightarrow \infty$) inequality~\eqref{eq_spohn} then yields the entropy production
\begin{equation}\label{eq_spohn_integrated_app}
   \Sigma=\Srel{\rho_0}{\rho_\mathrm{ss}}\geq0
\end{equation}
for the relaxation $\rho_0 \mapsto \rho_\mathrm{ss}$. Note that inequality~\eqref{eq_spohn} may not hold in the non-Markovian~\cite{erez2008thermodynamic} and/or strong-coupling regimes, where correlations or entanglement between the system and the bath may be appreciable~\cite{breuerbook}. By contrast, since the relative entropy is non-negative, Eq.~\eqref{eq_spohn_integrated_app} is valid for any system-bath coupling, not only in the Born-Markov regime~\cite{schloegl1966zur,deffner2011nonequilibrium}.

\par

For a Hamiltonian $H(t)$ that varies slowly compared to the thermalization time~\cite{alicki1979quantum}, the corresponding master equation is
\begin{equation}\label{eq_rho_dot_t}
  \dot\rho(t)=\mathcal{L}(t)\rho(t),
\end{equation}
where $\mathcal{L}(t)$ is the same Lindblad operator as in Eq.~\eqref{eq_app_master}, but with time-dependent coefficients (\cf Ref.~\cite{alicki1979quantum}), and an invariant state $\rho_\mathrm{ss}(t)$ that satisfies $\mathcal{L}(t)\rho_\mathrm{ss}(t)=0$. The generalization of inequality~\eqref{eq_spohn} to the case of~\eqref{eq_rho_dot_t} then 
leads, following integration, to 
\begin{equation}\label{eq_spohn_L_t}
  \Delta\mathcal{S}=\mathcal{S}(\rho_\mathrm{ss}(\infty))-\mathcal{S}(\rho_0) \geq - \kB\int_0^\infty\Tr\Big[\big(\mathcal{L}(t)\rho(t)\big)\ln\rho_\mathrm{ss}(t)\Big]\dd t.
\end{equation}

\par

If the time-dependent interaction is with a thermal bath at temperature $T$, then 
\begin{equation}\label{eq_app_rhoth_t}
  \rho_\mathrm{ss}(t)=\rho_\mathrm{th}(t)=\frac{1}{Z(t)}\exp\left(-\frac{H(t)}{\kB T}\right)
\end{equation}
is a thermal state for $H(t)$. Inequality~\eqref{eq_spohn_L_t} then yields
\begin{equation}\label{eq_spohn_t_integrated}
  \Delta\mathcal{S}\geq\frac{1}{T}\int_{0}^{\infty}\Tr\left[\dot{\rho}(t)H(t)\right]\dd t= \frac{\mathcal{E}_\mathrm{d}}{T}, 
\end{equation}
$\mathcal{E}_\mathrm{d}$ being the long-time limit $(t\rightarrow \infty)$ of the energy dissipated by the thermal bath.

\par 

We have pointed out~\cite{niedenzu2018quantum} that the second law (non-negative entropy production) may overestimate the actual system entropy change. This can be seen upon examining the relaxation of an initially non-passive state $\rho_0$ to a (passive) thermal state $\rho_\mathrm{th}$ through its interaction with a thermal bath at temperature $T$. According to the decomposition~\eqref{eq_DeltaEdiss_decomposition} of $\mathcal{E}_\mathrm{d}$, Eq.~\eqref{eq_spohn_t_integrated} can be written as
\begin{equation}\label{eq_DeltaS_QdT_DeltaW}
  \Delta\mathcal{S}\geq\frac{\mathcal{E}_\mathrm{d}}{T}=\frac{\mathcal{Q}+\left.\Delta\mathcal{W}\right|_\mathrm{d}}{T}.
\end{equation}
The second law thus reflects the entropy change bound to the total exchanged energy~\eqref{eq_def_DeltaEdiss}. However, as pointed out before, this energy can be decomposed into the two contributions~\eqref{eq_defs_heat_passive_work}, where only~\eqref{eq_def_heat} is associated to a change in entropy. Therefore, we contend that Eq.~\eqref{eq_DeltaS_QdT_DeltaW} ought to be replaced by a much tighter inequality that does not account for dissipated ergotropy, as explained below for different cases.

\subsection*{Entropy change in a thermal bath: Constant Hamiltonian}

Entropy is a state variable \cite{kondepudibook}, hence $\Delta\mathcal{S}=\mathcal{S}(\rho_\mathrm{th})-\mathcal{S}(\rho_0)$ is only determined by the initial state $\rho_0$ and the (passive) thermal steady state $\rho_\mathrm{th}$. Yet, Spohn's inequality~\eqref{eq_spohn_integrated_app} applied to  \textit{alternative evolution paths} from $\rho_0$ to $\rho_\mathrm{th}$, \textit{may give rise to different inequalities for the same} $\Delta\mathcal{S}$. In particular, we can choose a path that \textit{does not involve any dissipation of ergotropy} to the bath, i.e., it corresponds to first performing a unitary transformation to the passive state, $\rho_0\mapsto\pi_0$, then bringing this state in contact with the thermal bath, where it relaxes to the thermal steady-state $\rho_\mathrm{th}$. Inequality~\eqref{eq_spohn_integrated_app} applied on this path yields~\cite{niedenzu2018quantum}
\begin{equation}\label{eq_DeltaS_QdT}
  \Delta\mathcal{S}\geq\frac{\mathcal{Q}}{T},
\end{equation}
where $\mathcal{Q}$ is the same heat exchange as in Eq.~\eqref{eq_DeltaS_QdT_DeltaW}.

\par

In case of actual evolution, the system ergotropy decreases as a result of the relaxation caused by the thermal bath, from $\mathcal{W}_0\geq 0$, the ergotropy of the initial state $\rho_0$, down to vanishing ergotropy in the thermal steady state, so that 
\begin{equation}
\Delta\mathcal{W}|_\mathrm{d}=-\mathcal{W}_0\leq0.
\end{equation}
Hence, inequality~\eqref{eq_DeltaS_QdT} implies inequality~\eqref{eq_DeltaS_QdT_DeltaW}.

\subsection*{Entropy change of a system in a thermal bath: Time-dependent Hamiltonian}

As mentioned above, $\mathcal{Q}$ and $\Delta\mathcal{W}|_\mathrm{d}$ on the r.h.s.\ of Eq.~\eqref{eq_DeltaS_QdT_DeltaW} are \textit{path-dependent} if the Hamiltonian $H(t)$ slowly varies~\cite{alicki1979quantum}. This means that they depend not only on the initial state $\rho_0$ and the steady state $\rho_\mathrm{th}(\infty)$, which is a thermal state under the Hamiltonian $H(\infty)$. In particular, the path may be such that the time-dependent Hamiltonian generates a non-passive state at some point, even if the initial state is passive and the bath is thermal. Accordingly, we may choose a path void of \textit{initial ergotropy} by extracting the ergotropy of the initial state by a unitary process \textit{prior} to the interaction with the bath, thus resulting in the passive state $\pi_0$. If subsequently, this passive state is placed into contact with the thermal bath, it results in the steady state $\rho_\mathrm{th}(\infty)$. Our revised inequality for the latter step yields~\cite{niedenzu2018quantum}
\begin{equation}\label{eq_DeltaS_Qth}
  \Delta\mathcal{S}\geq\frac{\mathcal{Q}^\prime}{T},
\end{equation}
where
\begin{equation}\label{eq_def_Qth}
  \mathcal{Q}^\prime\coloneq\int_0^\infty\Tr[\dot\varrho(t)H(t)]\dd t
\end{equation}
is the heat exchanged with the bath along the chosen path. Here $\varrho(t)$ is the solution of the same thermal master equation that governs $\rho(t)$ but with the passive initial condition $\varrho_0=\pi_0$. For a constant Hamiltonian, Eq.~\eqref{eq_DeltaS_Qth} coincides with Eq.~\eqref{eq_DeltaS_QdT}.

\subsection*{Entropy change of a system in a non-thermal bath}\label{app_sigma_non-thermal}

We next consider the case of a system that interacts with a non-thermal bath while its Hamiltonian is kept constant. Due to this interaction, the system relaxes to the non-passive steady state
\begin{equation}\label{eq_rho_ss}
  \rho_\mathrm{ss}=U\pi_\mathrm{ss}U^\dagger.
\end{equation}
The entropy production~\eqref{eq_spohn_integrated_app} during the relaxation process $\rho_0\mapsto \rho_\mathrm{ss}$ then evaluates to
\begin{equation}\label{eq_sigma_non-thermal_bath}
  \Srel{\rho_0}{\rho_\mathrm{ss}}=\Srel{\tilde\rho_0}{\pi_\mathrm{ss}}\geq0,
\end{equation}
which equals the entropy production during the fictitious relaxation process $\tilde{\rho}_0\mapsto \pi_\mathrm{ss}$, where $\tilde\rho_0\coloneq U^\dagger\rho_0U.$

\par

In order to obtain the tightest inequality for the entropy change $\Delta\mathcal{S}$, we here, instead of~\eqref{eq_spohn_integrated_app}, propose the following relative-entropy inequality,
\begin{equation}\label{eq_srel}
  \Srel{\pi_0}{\pi_\mathrm{ss}}\geq 0.
\end{equation}
As before, the motivation for Eq.~\eqref{eq_srel} is that the entropy of any state $\rho$ is the same as its passive counterpart $\pi$. A general proof that Eq.~\eqref{eq_srel} is indeed a tighter inequality for $\Delta\mathcal{S}$ than Eq.~\eqref{eq_sigma_non-thermal_bath} can be found in Ref.~\cite{niedenzu2018quantum}.

\par

\begin{figure}
  \centering
  \includegraphics[width=0.50\columnwidth]{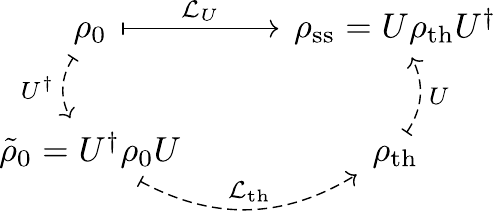}
  \caption{Two alternative evolution paths between $\rho_0$ and $\rho_\mathrm{ss}$. The underlying assumptions and the equivalence between the two paths are given in~\cite{niedenzu2018quantum}: The solid (physical) path corresponds to the interaction of the system with a non-thermal bath whereas the dashed path consists of two unitary transformations and the interaction with a thermal bath. Figure adapted from~\cite{niedenzu2018quantum}.}\label{fig_diagram}
\end{figure}

\par

We now consider the situation when the working fluid is a single-mode harmonic oscillator in a Gaussian state, i.e., an experimentally relevant~\cite{klaers2017squeezed} special case where the passive steady state is thermal, $\pi_\mathrm{ss}=\rho_\mathrm{th}$. As shown in Ref.~\cite{ekert1990canonical}, the system evolution (determined by a Markovian master equation~\cite{breuerbook} with Liouvillian $\mathcal{L}_U$) is then unitarily equivalent to the interaction of a transformed state $\tilde{\rho}(t)$ with a \emph{thermal} bath, which is described by the Liouvillian $\mathcal{L}_\mathrm{th}$ (Fig.~\ref{fig_diagram}). Inequality~\eqref{eq_sigma_non-thermal_bath} then yields
\begin{equation}\label{eq_sigma_rhotilde_thermal_bath}
  \Delta\mathcal{S}\geq\frac{\tilde{\mathcal{E}}_\mathrm{d}}{T},
\end{equation}
where $\tilde{\mathcal{E}}_\mathrm{d}$ is the change in the energy $\tilde E=\Tr[\tilde\rho H]$ of the unitarily transformed state $\tilde\rho$, as it relaxes from $\tilde\rho_0$ to $\rho_{\mathrm{th}}$ (second step in the dashed path in Fig.~\ref{fig_diagram}). By contrast, inequality~\eqref{eq_srel} keeps track of the initial ergotropy and evaluates to Eq.~\eqref{eq_DeltaS_QdT}.

\par

Hence, inequalities~\eqref{eq_DeltaS_QdT} and~\eqref{eq_DeltaS_Qth} also apply to the situation where the non-thermal bath relaxes the system to a state of the form $\rho_\mathrm{ss}=U\rho_\mathrm{th}U^\dagger$~\cite{niedenzu2018quantum}. The latter state naturally arises in machines of the first kind fuelled by a squeezed thermal bath~\cite{rossnagel2014nanoscale}.

\section{Efficiency bound of cyclic quantum engines powered by thermal or non-thermal baths}\label{sec_efficiency}

Based on the results of the previous section, we here show that reversible operation according to the second law does not yield a proper efficiency bound of cyclic engines fuelled by non-thermal baths whenever such baths impart both heat and ergotropy to the WF. This inadequacy, as discussed above, is due to the fact that the second law does not discern between dissipative changes in passive energy and ergotropy.

\par

\begin{figure}[h]
  \centering
  \includegraphics[width=0.6\columnwidth]{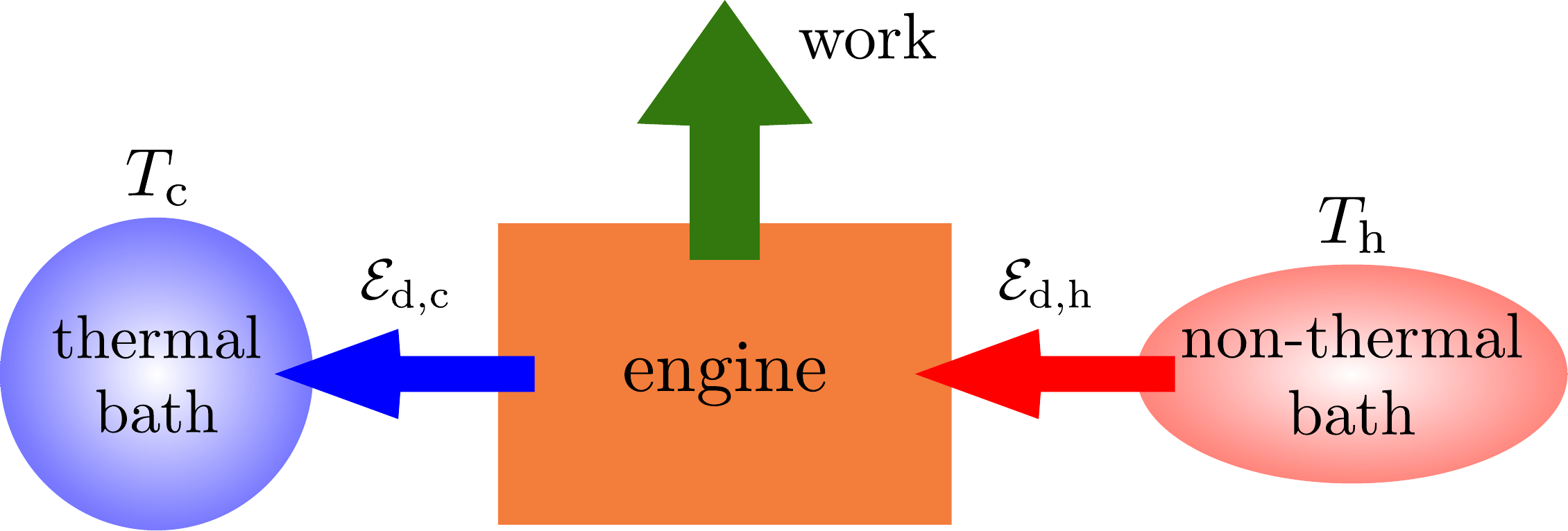}
  \caption{A quantum engine cyclically converts energy obtained from a non-thermal (e.g., squeezed thermal) bath into useful work that is extracted by a piston. In each cycle, the non-thermal bath provides the energy $\mathcal{E}_\mathrm{d,h}$. A fraction thereof is converted into work and an amount $\mathcal{E}_\mathrm{d,c}$ is dumped into the cold thermal bath. Figure adapted from~\cite{niedenzu2018quantum}.}\label{fig_engine}
\end{figure}

\par

Here we consider a broad class of cyclic quantum engines (Fig.~\ref{fig_engine}) that operate between a cold thermal bath (at temperature $T_\mathrm{c}$) and a hot (thermal or non-thermal) bath, while being subject to a \textit{time-dependent} drive (the ``piston''~\cite{alicki1979quantum}). The ``hot'' non-thermal bath is assumed to produce a non-passive state of the WF whose passive counterpart is thermal with temperature $T_\mathrm{h}>T_\mathrm{c}$. For a harmonic-oscillator (HO) WF interacting with a squeezed thermal bath~\cite{breuerbook,gardinerbook}, $T_\mathrm{h}$ is the temperature that the bath had before its squeezing. We allow the WF Hamiltonian to slowly change during the interaction with the baths~\cite{alicki1979quantum}. We only require that the WF attains its steady state by the end of the energizing stroke where it interacts with the hot  bath and the stroke where it interacts with the cold bath.

\par

The energizing stroke is described by a master equation~\cite{breuerbook} that evolves the WF to  a non-passive state $\rho_\mathrm{ss}(\infty)=U\rho_\mathrm{th}(\infty)U^\dagger$. To obtain the efficiency bound, we need to extract the ergotropy from the WF (by the piston via a suitable unitary transformation) before its interaction with the cold bath. In continuous cycles where both baths are simultaneously coupled to the WF~\cite{gelbwaser2013minimal,gelbwaser2015thermodynamics,mukherjee2016speed}, part of the ergotropy may then be dissipated into the cold bath, so that such cycles are inherently less efficient than stroke cycles. For time-dependent Hamiltonians, the requirement that $H(t)$ \textit{commutes with itself at all times}, e.g., a HO with time-independent eigenstates, will be adopted in this chapter for any interaction of a system with a bath, since Hamiltonians that do not commute with themselves at different times reduce the efficiency via ``quantum friction''~\cite{kosloff2013quantum,feldmann2006quantum}. 

\par

The entropy changes in the two strokes of the WF-bath interactions obey
\begin{subequations} 
\begin{align}
   \Delta\mathcal{S}_\mathrm{c} &\geq \frac{\mathcal{E}_\mathrm{d,c}}{T_\mathrm{c}}\\
   \Delta\mathcal{S}_\mathrm{h} &\geq \frac{\mathcal{Q}^\prime_\mathrm{h}}{T_\mathrm{h}}.
\end{align}
\end{subequations} 
Here $\mathcal{E}_\mathrm{d,c}\leq0$ is the change in the WF energy due to its interaction with the cold (thermal) bath and $\mathcal{Q}^\prime_\mathrm{h}\geq0$ is the heat that the WF would have received, if the the initial state was passive and the non-thermal bath were thermal (as in Eq.~\eqref{eq_def_Qth} and Fig.~\ref{fig_diagram}). The WF cyclically returns to its initial state, therefore $\Delta\mathcal{S}=\Delta\mathcal{S}_\mathrm{c}+\Delta\mathcal{S}_\mathrm{h}=0$ over a cycle. These two conditions yield the inequality
\begin{equation}\label{eq_condition_gen}
   \frac{\mathcal{E}_\mathrm{d,c}}{T_\mathrm{c}}+\frac{\mathcal{Q}^\prime_\mathrm{h}}{T_\mathrm{h}}\leq 0.
\end{equation}
Correspondingly, energy conservation over a cycle yields
\begin{equation}
  \mathcal{E}_\mathrm{d,c}+\mathcal{E}_\mathrm{d,h}+W=0,
\end{equation}
where $\mathcal{E}_\mathrm{d,h}$ is the dissipative energy change of the WF due to its interaction with the hot non-thermal bath (Fig.~\ref{fig_engine}). 

\par

The engine efficiency is defined as the ratio of the extracted work to the \textit{total energy} (heat and ergotropy) $\mathcal{E}_\mathrm{d,h}$
imparted by the hot (non-thermal or thermal) bath. Here we assume $\mathcal{E}_\mathrm{d,c}\leq0$ and $\mathcal{E}_\mathrm{d,h}\geq0$. The efficiency then reads
\begin{equation}\label{eq_app_eta}
  \eta\coloneq\frac{-W}{\mathcal{E}_\mathrm{d,h}}=1+\frac{\mathcal{E}_\mathrm{d,c}}{\mathcal{E}_\mathrm{d,h}}.
\end{equation}
From condition~\eqref{eq_condition_gen} it then follows that
  \begin{equation}
    \mathcal{E}_\mathrm{d,c}\leq -\frac{T_\mathrm{c}}{T_\mathrm{h}}\mathcal{Q}_\mathrm{d,h}^\prime,
  \end{equation}
which restricts the efficiency to
\begin{equation}\label{eq_etamax_gen}
  \eta\leq 1-\frac{T_\mathrm{c}}{T_\mathrm{h}}\frac{\mathcal{Q}^\prime_\mathrm{h}}{\mathcal{E}_\mathrm{d,h}}\eqcolon\eta_\mathrm{max}.
\end{equation}

\par

The efficiency bound~\eqref{eq_etamax_gen} does not only depend on the two temperatures~\cite{niedenzu2018quantum}. The bath characteristics (e.g., its squeezing parameter) determine the ratio $\mathcal{Q}^\prime_\mathrm{h}/\mathcal{E}_\mathrm{d,h}$ of the heat~\eqref{eq_def_Qth} to the total energy input from the hot bath. Eq.~\eqref{eq_etamax_gen} holds in the regime where the hot bath supplies energy and increases the WF entropy, $\mathcal{Q}^\prime_\mathrm{h} \geq 0$ and  $\mathcal{E}_\mathrm{d,h}>0$. The bound~\eqref{eq_etamax_gen} is then limited by unity, $\eta_\mathrm{max}\leq 1$. Unity efficiency is reached in the ``mechanical''-engine limit, $\mathcal{Q}^\prime_\mathrm{h}\rightarrow 0$, where the hot non-thermal bath only provides ergotropy. In the heat-engine limit, $\mathcal{Q}^\prime_\mathrm{h}\rightarrow \mathcal{E}_\mathrm{d,h}$, where only heat but no ergotropy is supplied by the hot bath, Eq.~\eqref{eq_etamax_gen} reduces to the Carnot bound $\eta_\mathrm{Carnot}=1-T_\mathrm{c}/T_\mathrm{h}$.

\par

By contrast, the bound $\eta_{\Sigma}$ that follows from the non-negativity of the entropy production may \textit{surpass $1$} which is unphysical (see Ref.~\cite{manzano2016entropy}). This can be seen from the second-law condition on entropy change over a cycle,
\begin{equation}
  \frac{\mathcal{E}_\mathrm{d,c}}{T_\mathrm{c}}+\frac{\tilde{\mathcal{E}}_\mathrm{d,h}}{T_\mathrm{h}}\leq0.
\end{equation}
Here $\tilde{\mathcal{E}}_\mathrm{d,h}$ is the energy change during the interaction with the thermal bath along the dashed path in Fig.~\ref{fig_diagram}, so that 
\begin{equation}\label{eq_eta_bound_spohn}
  \eta\leq 1-\frac{T_\mathrm{c}}{T_\mathrm{h}}\frac{\tilde{\mathcal{E}}_\mathrm{d,h}}{\mathcal{E}_\mathrm{d,h}}=:\eta_{\Sigma}.
\end{equation}
The bound in~\eqref{eq_eta_bound_spohn} exceeds $1$ if $\tilde{\mathcal{E}}_\mathrm{d,h}<0$, which occurs under \textit{excessive bath squeezing}, as explained below. The interaction with the effective thermal bath, which proceeds along the dashed path in Fig.~\ref{fig_diagram}, then \textit{reduces} the WF energy, whereas the interaction with the non-thermal bath along the solid path \textit{increases} the WF energy. By contrast, $\eta_\mathrm{max}$ in~\eqref{eq_etamax_gen} advocated by us does not exceed 1.

\par

The reason for the unphysicality of~\eqref{eq_eta_bound_spohn} becomes clear when the Hamiltonian is constant during the hot-bath stroke. Then 
\begin{equation}
  \tilde{\mathcal{E}}_\mathrm{d,h}=\widetilde{\mathcal{Q}}+\widetilde{\Delta\mathcal{W}|_\mathrm{d}},
\end{equation}
where $\widetilde{\Delta\mathcal{W}|_\mathrm{d}}\leq 0$ is the ergotropy lost to the effective thermal bath in the second stage of the dashed path (Fig.~\ref{fig_diagram}). The more we squeeze the non-thermal bath, the more ergotropy is lost to the effective thermal bath, until the regime $\tilde{\mathcal{E}}_\mathrm{d,h}<0$ is attained, causing Eq.~\eqref{eq_eta_bound_spohn} to exceed 1. 

\par 

There exists a regime~\cite{niedenzu2016operation}, wherein such a machine of the first kind acts simultaneously as an engine and a refrigerator that cools the cold bath. In this regime $\mathcal{E}_\mathrm{d,c}>0$  and $\mathcal{Q}_\mathrm{d,h}^\prime<0$. Then also the cold bath provides energy to the WF and the efficiency becomes
\begin{equation}\label{eq_app_eta_1}
  \eta=\frac{-W}{\mathcal{E}_\mathrm{d,h}+\mathcal{E}_\mathrm{d,c}}=\frac{\mathcal{E}_\mathrm{d,h}+\mathcal{E}_\mathrm{d,c}}{\mathcal{E}_\mathrm{d,h}+\mathcal{E}_\mathrm{d,c}}=1.
\end{equation}
Namely, it is not restricted by any condition on $\Delta\mathcal{S}$ and hence by the second law.

\section{Stroke-based quantum engines powered by non-thermal baths}

\subsection*{Machines of the first kind}

We now consider explicit examples of stroke-based (``reciprocal'') bath-powered engine cycles which are not restricted by the second law, but by other constraints on their entropy. These cycles conform to the classification of the engine as a \textit{machine of the first kind} in the introduction.

\subsubsection{Modified Otto cycle powered by a squeezed thermal bath}

\begin{figure*}[h]
  \centering
  \includegraphics[width=0.6\columnwidth]{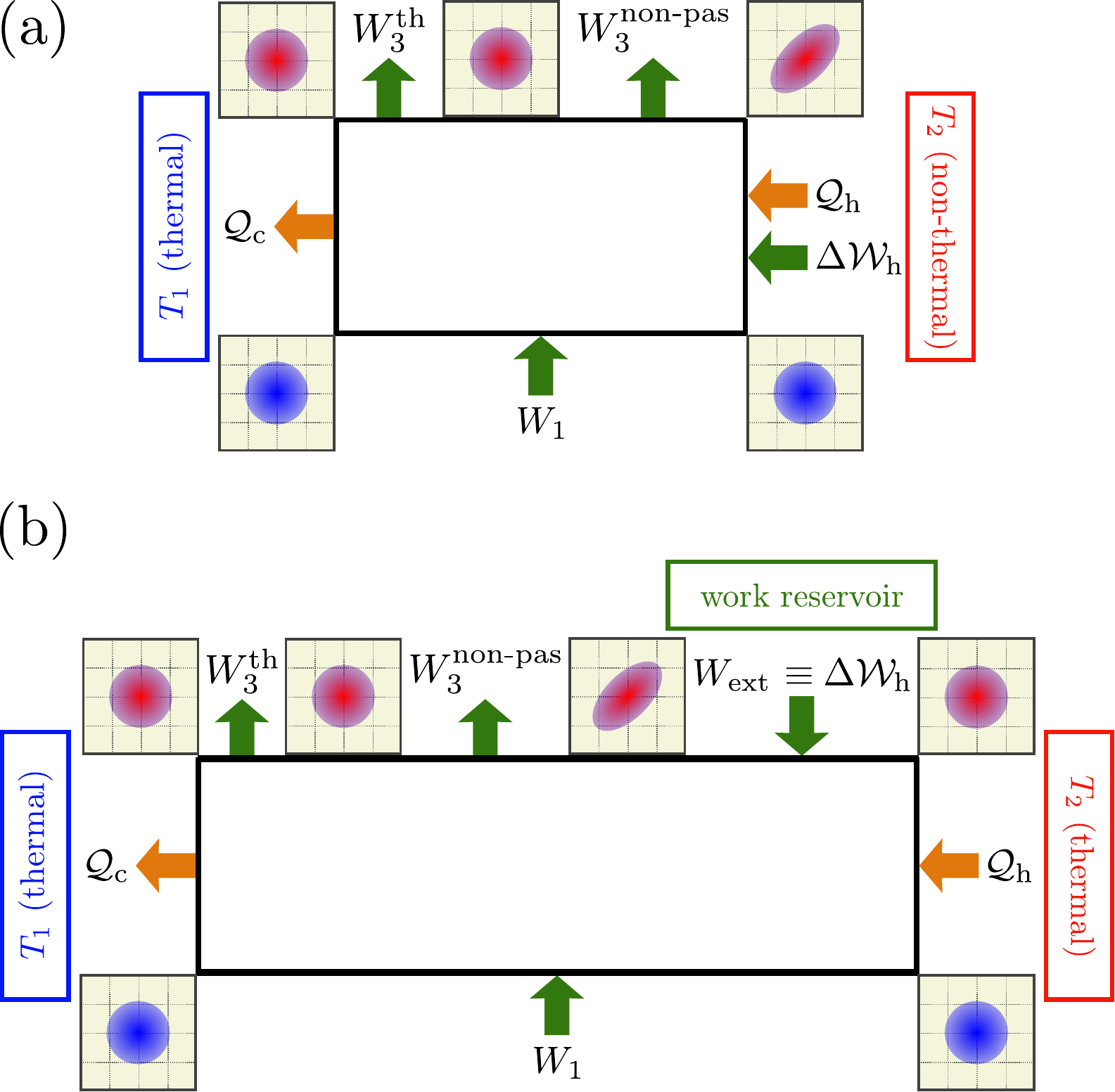}
  \caption{(a)~Modified Otto cycle for an oscillator WF and a squeezed thermal bath: In the first stroke the work $W_1$ is invested by the piston to compress the (thermal) WF. The second stroke consists of the interaction of the WF with a squeezed thermal bath, from which passive energy $\mathcal{Q}_\mathrm{h}$ and ergotropy $\Delta\mathcal{W}_\mathrm{h}$ is imparted to the WF. As a modification of the standard Otto cycle, now a unitary transformation is performed on the WF to extract its ergotropy in the form of work $W_3^\text{non-pas}$. The WF is now passive (thermal) and the work $W_3^\mathrm{th}$ is extracted by the piston by expanding the WF, as in the standard Otto cycle. Finally, the cycle is closed by the interaction of the WF with a cold thermal bath at temperature $T_\mathrm{c}$ to which the heat $\mathcal{Q}_\mathrm{c}$ is released. (b)~Equivalent hybrid machine yielding the same work and efficiency as the modified Otto cycle in~(a): Instead of being fuelled by a squeezed thermal bath, the WF first interacts with a hot thermal bath at temperature $T_\mathrm{h}$ that provides the heat $\mathcal{Q}_\mathrm{c}$. Afterwards, an external work reservoir (e.g., a battery) investes the work $W_\mathrm{ext}$ to squeeze the WF. This increases the WF ergotropy by the same amount $\Delta\mathcal{W}_\mathrm{h}$ that the squeezed thermal bath produced in~(a). This demonstrates the hybrid character of machines of the first kind. Figure adapted from~\cite{niedenzu2016operation}.}\label{fig_quantum_otto_cycle_2_new}
\end{figure*}

\par 

We first consider machines powered by a squeezed thermal bath that obey the modified Otto cycle proposed in~\cite{niedenzu2016operation}. In addition to the two isentropic strokes (adiabatic compression and decompression of the WF) and the two isochoric strokes (interaction with the baths under a fixed Hamiltonian) of the standard quantum Otto cycle~\cite{kosloff2017quantum}, we add one additional \textit{ergotropy-extraction stroke}. The latter may be implemented by abruptly ramping up the HO frequency and then gradually ramping this frequency down~\cite{graham1987squeezing,agarwal1991exact,averbukh1994enhanced}. This cycle is schematically shown in Fig.~\ref{fig_quantum_otto_cycle_2_new}a.

\par

As the interaction of the WF with the bath is isochoric, we have $\mathcal{Q}^\prime_\mathrm{h}=\Delta E_\mathrm{pas,h}$ and $\mathcal{E}_\mathrm{d,h}=\Delta E_\mathrm{pas,h}+\Delta \mathcal{W}_\mathrm{h}$, where $\Delta E_\mathrm{pas,h}$
is the passive energy change and $\Delta \mathcal{W}_\mathrm{h}$ is the ergotropy change, respectively. The efficiency bound~\eqref{eq_etamax_gen} of this Otto-like cycle then reads
\begin{equation}\label{eq_etamax_otto}
  \eta_\mathrm{max}=1-\frac{T_\mathrm{c}}{T_\mathrm{h}}\frac{\Delta E_\mathrm{pas,h}}{\Delta E_\mathrm{pas,h}+\Delta \mathcal{W}_\mathrm{h}}.
\end{equation}
In general, any engine cycle wherein the interaction with the hot bath is isochoric and sufficiently long (for the WF to reach steady state) abides by the bound~\eqref{eq_etamax_otto}.

\par 

The interaction of the WF with the squeezed thermal bath is governed by the master equation~\cite{gardinerbook}
\begin{equation}\label{eq_app_master_squeezing}
  \dot\rho=\kappa(N+1)\mathcal{D}(a,a^\dagger)[\rho]+\kappa N\mathcal{D}(a^\dagger,a)[\rho]-\kappa M\mathcal{D}(a,a)[\rho]-\kappa M\mathcal{D}(a^\dagger,a^\dagger)[\rho].
\end{equation}
Here the dissipator is defined as $\mathcal{D}(A,B)[\rho]\coloneq 2A\rho B-BA\rho-\rho BA$, $a$ and $a^{\dagger}$ are the WF creation and annihilation operators, $\kappa$ denotes the decay rate and (w.l.o.g. the squeezing phase is zero)
\begin{subequations}
  \begin{align}\label{eq_master_squeezing_standard_coefficients}
    N&\coloneq\bar{n}_\mathrm{h}(\cosh^2r+\sinh^2r)+\sinh^2r\\
    M&\coloneq-\cosh r\sinh r (2\bar{n}_\mathrm{h}+1),
  \end{align}
\end{subequations}
$\bar{n}_\mathrm{h}=[\exp(\hbar\omega/[\kB T_\mathrm{h}])-1]^{-1}$ being the thermal excitation number of the bath at the WF-oscillator frequency $\omega$ and $r$ the squeezing parameter. 

\par 

\begin{figure}[h]
  \centering
  \includegraphics[width=0.5\columnwidth]{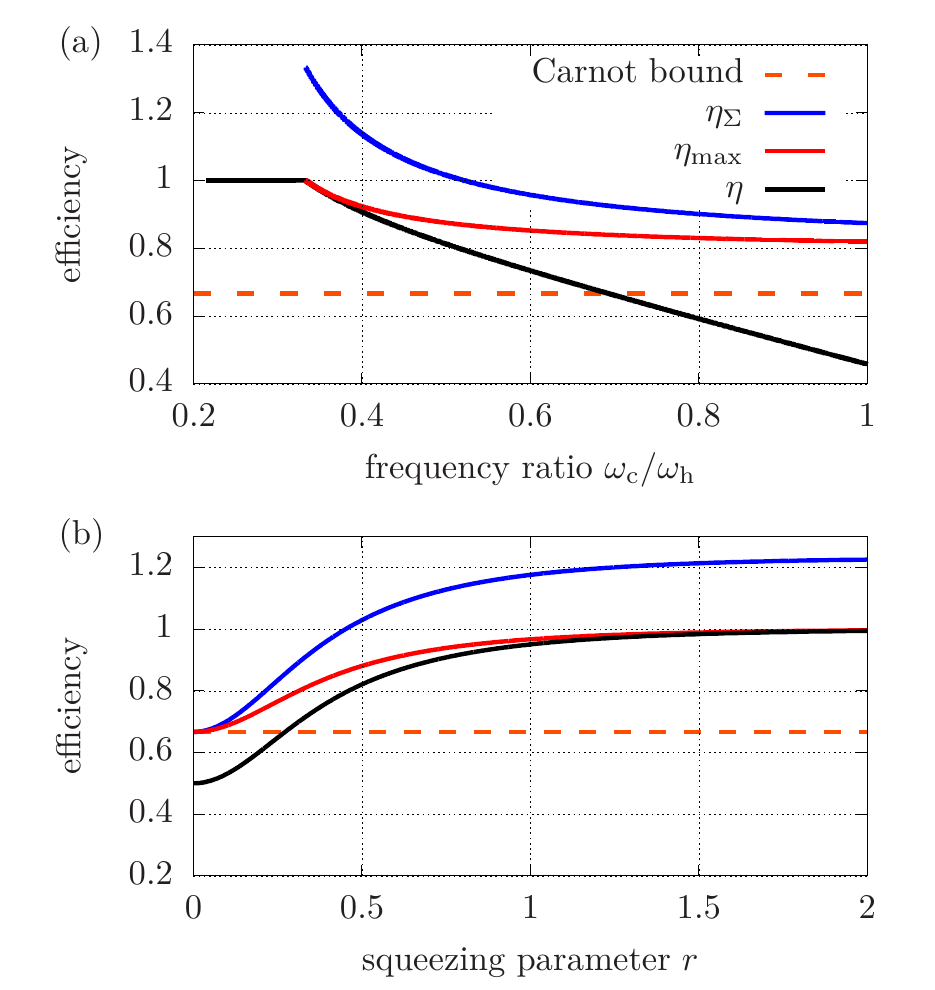}
  \caption{Efficiency~\eqref{eq_the-efficiency} of the modified Otto cycle as a function of (a)~the frequency ratio and~(b) the squeezing parameter. The second-law bound $\eta_\Sigma$ [Eq.~\eqref{eq_eta-sigma-otto}] does not limit the efficiency and may even surpass~$1$. By contrast, the bound~\eqref{eq_eta-max-otto} interpolates between the Carnot bound (in the heat-engine limit) and unity (in the mechanical-engine limit). Parameters: $T_\mathrm{h}=3T_\mathrm{c}$ and (a)~$r=0.5$ and (b)~$\omega_\mathrm{c}=\omega_\mathrm{h}/2$. Figure adapted from~\cite{niedenzu2018quantum}.}\label{fig_efficiency_otto}
\end{figure}

The steady state of Eq.~\eqref{eq_app_master_squeezing} is a squeezed thermal state of the WF. The deviation of the squeezed WF's excitation number from thermal equilibrium is~\cite{gardinerbook}
\begin{equation}
  \Delta\nbar_\mathrm{h}=(2\nbar_\mathrm{h}+1)\sinh^2(r)>0.
\end{equation}
The efficiency bounds $\eta_{\Sigma}$ [Eq.~\eqref{eq_eta_bound_spohn}] and $\eta_\mathrm{max}$ [Eq.~\eqref{eq_etamax_otto}] then evaluate to~\cite{niedenzu2018quantum}
\begin{equation}\label{eq_eta-sigma-otto}
  \eta_{\Sigma}=1-\frac{T_\mathrm{c}}{T_\mathrm{h}}\frac{\nbar_\mathrm{h}-\nbar_\mathrm{c}-\Delta\nbar_\mathrm{c}}{\nbar_\mathrm{h}+\Delta\nbar_\mathrm{h}-\nbar_\mathrm{c}}
\end{equation}
and
\begin{equation}\label{eq_eta-max-otto}
  \eta_\mathrm{max}=1-\frac{T_\mathrm{c}}{T_\mathrm{h}}\frac{\nbar_\mathrm{h}-\nbar_\mathrm{c}}{\nbar_\mathrm{h}+\Delta\nbar_\mathrm{h}-\nbar_\mathrm{c}}.
\end{equation}
The efficiency of the modified Otto cycle evaluates to~\cite{niedenzu2016operation}
\begin{equation}\label{eq_the-efficiency}
  \eta=1-\frac{(\nbar_\mathrm{h}-\nbar_\mathrm{c})\omega_\mathrm{c}}{(\nbar_\mathrm{h}+\Delta\nbar_\mathrm{h}-\nbar_\mathrm{c})\omega_\mathrm{h}}.
\end{equation}
Equation~\eqref{eq_the-efficiency} holds for $\mathcal{E}_\mathrm{d,c}\leq0$; for $\mathcal{E}_\mathrm{d,c}\geq0$ the machine operates as an engine and a refrigerator for the cold bath with $\eta=1$ [Eq.~\eqref{eq_app_eta_1}]. The machine acts as an engine for $\mathcal{E}_\mathrm{d,h}\geq0$, i.e., it delivers work for $\nbar_\mathrm{h}+\Delta\nbar_\mathrm{h}\geq\nbar_\mathrm{c}$. In Fig.~\ref{fig_efficiency_otto} we compare $\eta_\Sigma$ [Eq.~\eqref{eq_eta-sigma-otto}] obtained from the reversibility condition with the physical bound $\eta_\mathrm{max}$ [Eq.~\eqref{eq_eta-max-otto}] discussed above.

\par 

\begin{figure}[h]
  \centering
  \includegraphics[width=0.5\columnwidth]{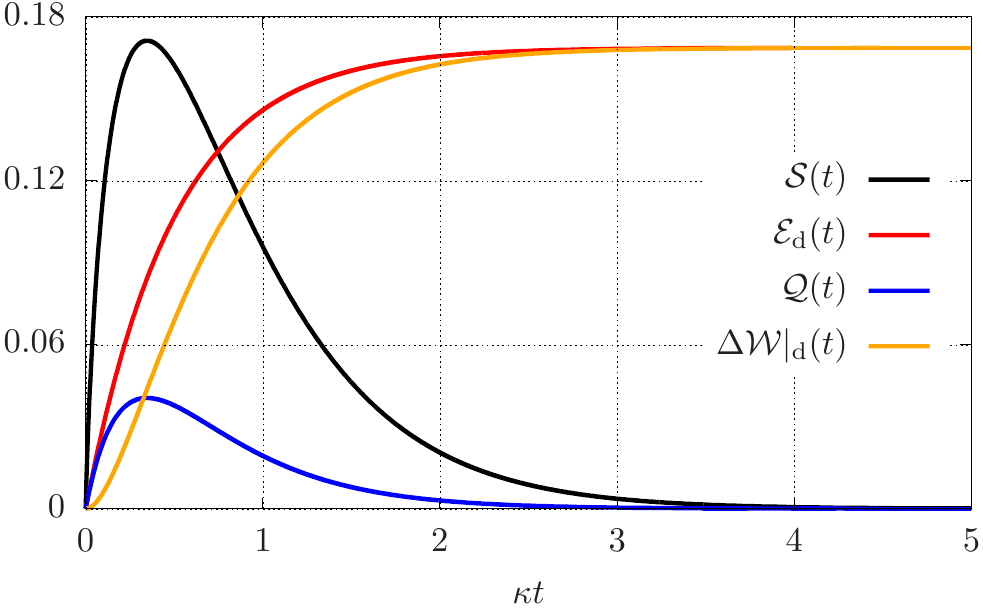}
  \caption{Entropy, energy, passive-energy, and ergotropy changes for a harmonic oscillator initialized in the vacuum state that interacts with a squeezed-vacuum bath according to the master equation~\eqref{eq_app_master_squeezing}. Figure adapted from~\cite{niedenzu2018quantum}.}\label{fig_squeezed_cavity}
\end{figure}

The difference between a standard heat engine (which, by definition, is energized exclusively by heat) and this hybrid (thermo-mechanical) \textit{machine of the first kind} becomes apparent in the extreme case $T_\mathrm{c}=T_\mathrm{h}=0$ ($\nbar_\mathrm{c}=\nbar_\mathrm{h}=0$). A machine of the first kind can then still deliver work,
\begin{equation}\label{eq_work_zero_temperature}
  W=-\hbar(\omega_\mathrm{h}-\omega_\mathrm{c})\Delta\nbar_\mathrm{h}<0,
\end{equation}
although no heat is imparted to the WF by the (pure-state) bath. The machine is then an \emph{effectively mechanical} engine, energized by $\mathcal{E}_{\mathrm{d,h}}=\hbar\omega_\mathrm{h}\Delta\nbar_\mathrm{h}>0$, which is pure ergotropy transfer from the zero-temperature bath. The WF energy increase in the second stroke is then \emph{effectively isentropic} and does not involve any net heat exchange with the bath (Fig.~\ref{fig_squeezed_cavity}).

\subsubsection{An equivalent hybrid cycle}

The above modified Otto cycle (Fig.~\ref{fig_quantum_otto_cycle_2_new}a) may be replaced by an equivalent cycle (Fig.~\ref{fig_quantum_otto_cycle_2_new}b) involving a hot thermal bath at temperature $T_\mathrm{h}$ and an external work source (which is \textit{not the piston}) \cite{niedenzu2016operation}. After the second stroke this external work source performs a unitary transformation on the thermal WF state that transforms it into the same non-passive state that it would become via contact with a non-thermal bath. The amount of work invested by this device is the same ergotropy of the WF state that a non-thermal bath would provide. This equivalent cycle demonstrates the \emph{hybrid thermo-mechanical} nature of the engine. The equivalence of this cycle and the modified Otto cycle follows from our analysis, whereby the heat provided by the non-thermal bath is the \emph{same} as the heat that a thermal bath at temperature $T_\mathrm{h}$ would have provided [cf. Eq.~\eqref{eq_condition_gen}]. The energy surplus imparted by the non-thermal bath was identified to be the work obtained from ergotropy. This equivalence supports our conclusion that the maximum efficiency of this cycle is not a thermodynamic bound, since the work imparted by the auxiliary work reservoir is not bounded by the second law of thermodynamics, whereas the heat exchanges are.

\subsubsection{Modified Carnot cycle powered by a squeezed thermal bath}

\begin{figure}[h]
  \centering
  \includegraphics[width=0.5\columnwidth]{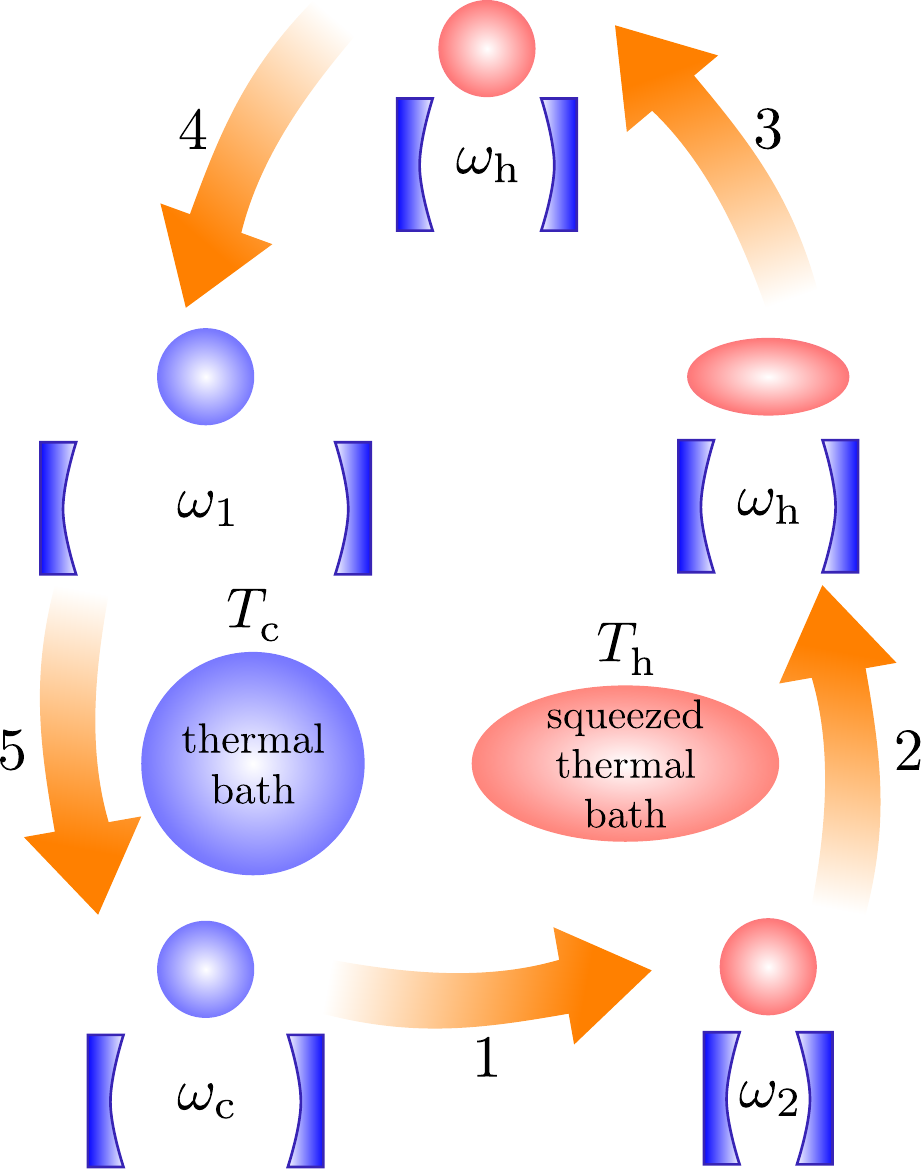}
  \caption{The modified Carnot cycle starts with a thermal state with frequency $\omega_\mathrm{c}$ and temperature $T_\mathrm{c}$ (lower left corner). In stroke~$1$, the mode undergoes an adiabatic compression to frequency $\omega_2=\omega_\mathrm{c}T_\mathrm{h}/T_\mathrm{c}$ and temperature $T_\mathrm{h}>T_\mathrm{c}$. Thereafter, in the energizing stroke~$2$, the frequency is slowly reduced to $\omega_\mathrm{h}\leq\omega_2$ while the mode is connected to the squeezed thermal bath, yielding a squeezed thermal steady state. Its ergotropy is extracted in stroke~$3$ by an ``unsqueezing'' unitary operation, resulting in a thermal state with temperature $T_\mathrm{h}$. In stroke~$4$, the frequency is again adiabatically reduced to $\omega_1=\omega_\mathrm{h}T_\mathrm{c}/T_\mathrm{h}$ such that the mode attains the temperature $T_\mathrm{c}$. Finally, stroke~$5$ is an isothermal compression back to the initial state. Figure adapated from~\cite{niedenzu2018quantum}.}\label{fig_carnot}
\end{figure}

\begin{figure}[h]
  \centering
  \includegraphics[width=0.5\columnwidth]{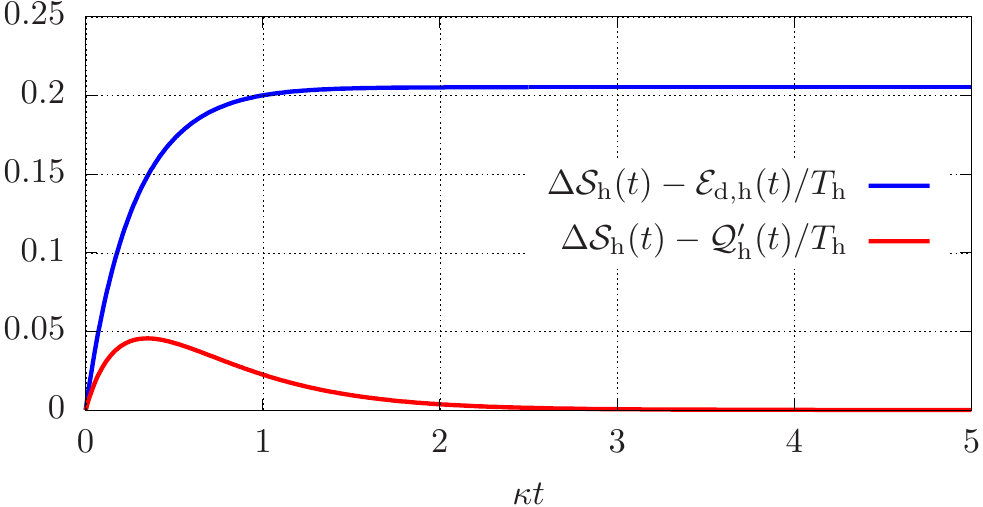}
  \caption{Change in entropy (in units of $\kB$) during stroke~$2$ of the modified Carnot cycle in Fig.~\ref{fig_carnot} as a function of the stroke duration obtained by a numerical integration of the master equation. The upper (blue) curve corresponds to the second-law inequality~\eqref{eq_spohn_t_integrated} which is far from being saturated. By contrast, our proposed inequality~\eqref{eq_DeltaS_Qth} is saturated (i.e., the equality sign applies) for sufficiently long stroke duration (red lower curve); here $\Delta\mathcal{S}_\mathrm{h}(t)=\mathcal{S}(\rho(t))-\mathcal{S}(\rho_0)$. Parameters: $\omega(t)=(25-0.05\kappa t)\kappa$, $\kB T_\mathrm{h}=5\hbar\kappa$ and $r=0.2$, $\kappa$ being the decay rate of the cavity-mode WF. Figure adapated from \cite{niedenzu2018quantum}.}\label{fig_carnot_engine}
\end{figure}

We finally consider a cycle powered by a squeezed thermal bath that comprises the four strokes of the ordinary thermal Carnot cycle~\cite{carnotbook,kondepudibook} and an additional ergotropy-extraction stroke (stroke $3$ in Fig.~\ref{fig_carnot}). 
Stroke $2$ is isothermal expansion wherein the state $\varrho(t)$ is always in thermal equilibrium. Hence, from Eq.~\eqref{eq_DeltaS_Qth} we have $\mathcal{Q}^\prime_\mathrm{h}=T_\mathrm{h}\Delta\mathcal{S}_\mathrm{h}$ (Fig.~\ref{fig_carnot_engine}). Stroke $5$ is isothermal compression, i.e., $\mathcal{E}_\mathrm{d,c}=T_\mathrm{c}\Delta\mathcal{S}_\mathrm{c}$. The condition of vanishing entropy change over a cycle corresponds to the equality sign in condition~\eqref{eq_condition_gen}. Hence, the efficiency of this cycle is the bound $\eta_{\mathrm{max}}$ in Eq.~\eqref{eq_etamax_gen}.

\par

\par

We conclude that the bound $\eta_{\mathrm{max}}$ is lower than $\eta_\mathrm{\Sigma}$~\cite{niedenzu2018quantum} for all possible engine cycles that contain a ``Carnot-like'' energizing stroke, namely, a stroke characterized by a slowly-changing Hamiltonian and an initial thermal state at temperature $T_\mathrm{h}$, such that $\mathcal{Q}^\prime_\mathrm{h}=T_\mathrm{h}\Delta\mathcal{S}_\mathrm{h}$. The Carnot-like cycle always operates at the maximum efficiency~\eqref{eq_etamax_gen}, even when both passive thermal energy and ergotropy are imparted by this bath.

\subsection*{Machines of the second kind}\label{sec_thermal_otto}

There are machines wherein the WF does not draw both work and heat from the non-thermal bath, but is instead thermalized by this bath, in spite of the bath being non-thermal. Then, the excitation $\nbar_\mathrm{h}+\Delta\nbar_\mathrm{h}$ corresponds to a \emph{real} temperature $\Treal$ (the WF relaxes to a thermal state with this temperature if left in contact with the non-thermal bath) of the WF~\cite{alicki2015nonequilibrium}, such that \cite{niedenzu2016operation}
\begin{subequations}\label{eq_first_second_law_thermal_wf}
\begin{equation}\label{eq_first_second_law_thermal_wf_1}
  \mathcal{Q}_\mathrm{c}+\mathcal{Q}_\mathrm{h}+W=0
\end{equation}
and
\begin{equation}\label{eq_first_second_law_thermal_wf_2}
  \frac{\mathcal{Q}_\mathrm{h}}{\Treal}+\frac{\mathcal{Q}_\mathrm{c}}{T_\mathrm{c}}\leq 0.
\end{equation}
\end{subequations}

\par

Thus, in this regime the machine operates as a genuine heat engine whose efficiency is restricted by the Carnot bound
\begin{equation}\label{eq_carnot_thermal_wf}
  \eta=\frac{-W}{\mathcal{Q}_\mathrm{h}}\leq1-\frac{T_\mathrm{c}}{\Treal}\equiv\eta_\mathrm{Carnot}
\end{equation}
corresponding to this real temperature. The (``original'') temperature $T_\mathrm{h}$ of the bath, prior to its transformation into a non-thermal state, plays no role; the only temperatures that matter are $T_\mathrm{c}$ and $\Treal$. These temperatures appear in~\eqref{eq_first_second_law_thermal_wf_2}, which, together with the first law~\eqref{eq_first_second_law_thermal_wf_1}, gives rise to the Carnot bound~\eqref{eq_carnot_thermal_wf}.

\par

For a HO WF, such a machine of the second kind can be realized, e.g., for a cavity being fuelled by a phaseonium bath where $\Treal=T_\varphi$~\cite{scully2003extracting}, by its $N$-level generalization~\cite{turkpence2016quantum}, or by a beam of entangled atom dimers~\cite{dag2018temperature}. The non-thermal character of the bath presents an advantage to engine operation only if $\Treal>T_\mathrm{h}$, which corresponds to $\Delta\nbar_\mathrm{h}>0$. Contrary to machines of the first kind, in machines of the second kind $\Delta\nbar_\mathrm{h}$ may, in principle, become negative. This case is exemplified by a phaseonium bath with the wrong choice of phase~$\varphi$~\cite{scully2003extracting}.

\section{Catalysis of heat-to-work conversion in quantum machines}

We have recently shown~\cite{ghosh2017catalysis} that under non-linear (quadratic) pumping the piston mode of a quantum heat engine evolves into a squeezed thermal state that strongly enhances its work capacity (ergotropy) compared to its linearly-pumped or unpumped counterparts. The resulting effects are that the output power and efficiency of heat-to-work conversion are drastically enhanced. Yet, since the engine is fueled by thermal baths, its efficiency is limited by the Carnot bound~\cite{kondepudibook}. In describing the enhanced performance of the engine we resort to the concept of \textit{catalysis}, whereby a small amount of catalyst (here a weak pump) strongly enhances the heat-to-work conversion. 

\par 

\subsection*{Model of catalyzed quantum heat engine}
\begin{figure}[h]
	\centering
		\includegraphics[width=\columnwidth]{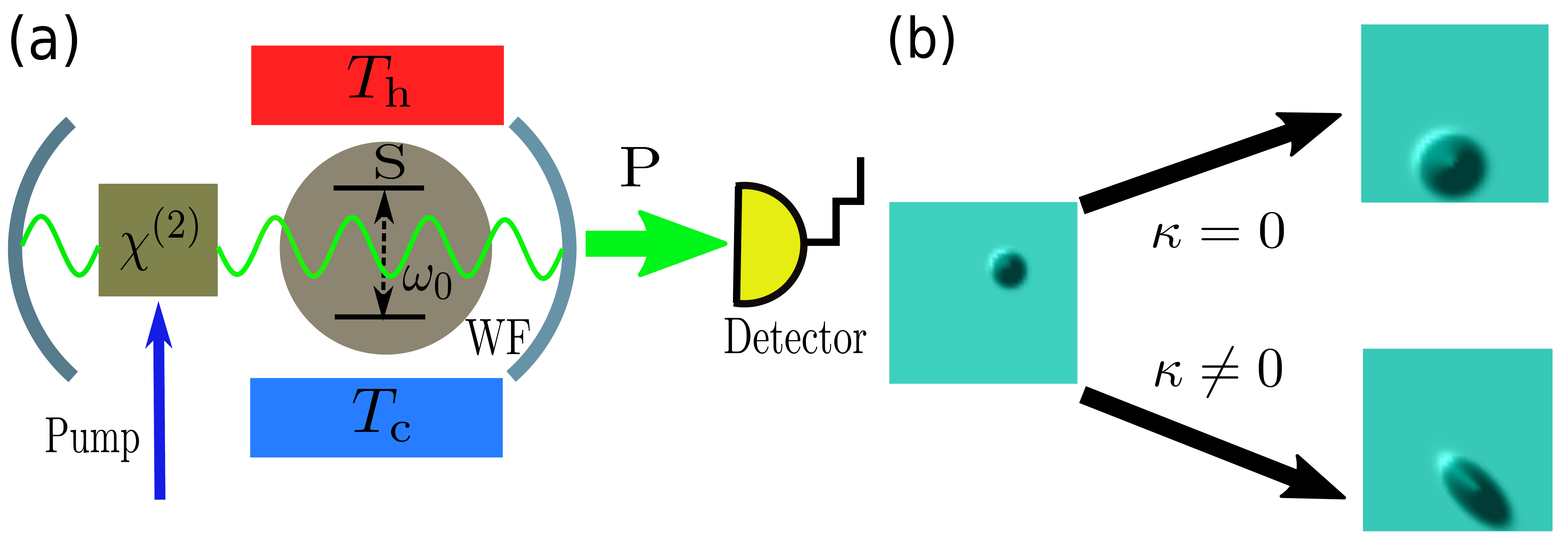}
\caption{(a)~A schematic diagram of a catalyzed quantum heat engine: The two-level WF, S is continuously coupled to cold and hot thermal baths and to a non-linearly-pumped piston mode P. (b)~Evolution of the Wigner phase-space distribution function of an initial coherent state in the presence ($\kappa \neq 0$) and absence ($\kappa = 0$) of a non-linear pump.}
	\label{fig:14}
\end{figure}

Our model of a catalyzed quantum heat engine is shown in Fig.~\ref{fig:14}a. The Hamiltonian has the form ($\hbar = 1$)
\begin{gather}
H_{\mathrm{tot}}=\sum_{j=\mathrm{h,c}} (H^j_{\mathrm{SB}}+H^j_{\mathrm{B}})+H_{\mathrm{S+P}}+H_{\mathrm{pump}}(t),
\label{eq:hsp}
\end{gather}
where, $H^j_{\mathrm{B}}$ is the free Hamiltonian of the cold (c) and hot (h) baths and $H^j_{\mathrm{SB}}=\sigma_{\mathrm{X}}B_j$ is their coupling to the WF. The S-P interaction is described by
\begin{gather}
H_{\mathrm{S+P}}=H_\mathrm{S}+H_\mathrm{P}+H_{\mathrm{SP}}, \label{eq:htot}\nonumber\\
H_\mathrm{S}=\frac{1}{2}\omega_{0}\sigma_{Z};\;H_{\mathrm{P}}=\nu a^{\dagger}a;\; H_{\mathrm{SP}}=g\sigma_{Z}\otimes(a+a^{\dagger}).
\label{eq:h-s+p}
\end{gather}
The P mode is isolated from the baths. Yet, the latter still modify its energy and entropy (Fig.~\ref{fig:14}) indirectly via S. This implies that the state of the piston cannot be fully cyclic, it must inevitably keep changing.

\par 

The catalysis is induced by the coupling of P to a (degenerate) parametric amplifier~\cite{gardinerbook} via the Hamiltonian
\begin{gather}
H_{\mathrm{pump}}(t)=\frac{i}{2}\kappa e^{-2i\nu t}{a^{\dag}}^{2}+\mathrm{h.c}, \label{eq:sqz}
\end{gather}
where $|\kappa|\;(0 < |\kappa| \leq 1)$ is the undepleted pumping rate, taken to be real. The quadratic form of $H_{\mathrm{pump}}(t)$ generates squeezing~\cite{gardinerbook}. We show that two \textit{not-additive} processes may reinforce each other, thereby producing the catalysis: The enhancement of the ergotropy of P and the amplification of P by the S-B coupling.

\par 

The present model is realizable by a cavity-based non-linear parametric amplifier~\cite{gardinerbook} that couples to two heat baths with different temperatures and spectra. S can be a superconducting flux qubit that is dispersively coupled to P which can be realized by a phonon mode of a nano-mechanical cantilever~\cite{aspelmeyer2014cavity} or by a field mode~\cite{ghosh2018two-level}. In either realization, the quantized quadrature $a+a^{\dag}$ of the P-mode acts on the flux qubit energy $\sigma_Z$~\cite{gelbwaser2013minimal}.

\subsection*{Model analysis: Derivation of the Lindblad master equation}
In order to derive a Lindblad master equation, we first diagonalize the Hamiltonian~\eqref{eq:h-s+p} by the unitary transformation 
\begin{gather}
a\mapsto b=\mathcal{U}^{\dagger}a\mathcal{U},\;\; \sigma_{k}\mapsto\widetilde{\sigma}_{k}=\mathcal{U}^{\dagger}\sigma_{k}\mathcal{U};\;\;
\mathcal{U}=\exp\left[{\frac{g}{\nu}(a^{\dag}-a)\sigma_{Z}}\right];\;\; (k=X,Y,Z).\label{dress}
\end{gather}
In terms of these new operators, the evolution of the piston mode is found to be~\cite{ghosh2017catalysis}
\begin{gather}
\dot{\rho}_\mathrm{P} = \frac{\Gamma +D}{2}\bigl([b, \rho_\mathrm{P} b^{\dagger}] + [b \rho_\mathrm{P}, b^{\dagger}] \bigr) + \frac{D}{2}\bigl([b^{\dagger}, \rho_\mathrm{P} b] + [b^{\dagger} \rho_\mathrm{P}, b] \bigr)
+\frac{\kappa}{2}[{b^{\dag}}^{2},\rho_\mathrm{P}(t)]-\frac{\kappa^{*}}{2}[b^{2},\rho_\mathrm{P}(t)].
\label{ME_reduced}
\end{gather}
Here both the drift (gain or amplification) and diffusion (thermalization) rates 
\begin{gather}
\Gamma= \left(\frac{g}{\nu} \right)^{2}\Big((G(\omega_+)-G(\omega_-))\rho_{11}\nonumber
+(G(-\omega_{-})-G(-\omega_{+}))\rho_{00} \Big);\nn \\
 D=\left(\frac{g}{\nu} \right)^{2}\left(\left(G(\omega_-)\rho_{11}+G(-\omega_{+})\rho_{00}\right)\right),
\label{eq:gamad}
\end{gather}
depend on the sum of the baths' response spectra $G(\omega)=\sum_{j=\mathrm{h,c}}G_{j}(\omega)$ evaluated at frequencies $\omega_{\pm}=\omega_{0}\pm \nu$. They satisfy the Kubo-Martin Schwinger (KMS) condition~\cite{breuerbook} $G_{j}(\omega)=e^{\omega/T_{j}}G_{j}(-\omega)$, and $\rho_{11}$, $\rho_{00}$ are the  populations of the upper and lower states of S. One may verify that even with the original operators (cf. Eq.~\eqref{dress}), the master equation for P to second order in $g/\nu$ does not change the qualitative nature of our results~\cite{ghosh2017catalysis}. Hence, in the following we consider Eq.~\eqref{ME_reduced} as our starting equation.

\par

For appropriate values of $G_j(\omega_{\pm})$ one can achieve gain, i.e., $\Gamma < 0$. The following choice is optimal for a heat engine~\cite{gelbwaser2013minimal,gelbwaser2014heat}: $G_\mathrm{h}(\omega_0-\nu)=G_\mathrm{c}(\omega_0-\nu)=0,\;G_\mathrm{h}(\omega_0+\nu) \gg G_\mathrm{c}(\omega_0+\nu),\;G_\mathrm{c}(\omega_0)\gg G_\mathrm{h}(\omega_0)$. Namely, the spectral densities of the baths nearly non-overlapping, the cold bath spectral density is centered around $\omega_0$ and the hot bath spectral density around $\omega_0+\nu$. The sum $D+\Gamma \ge 0$, so that $\Gamma < 0$, implies $D/|\Gamma|\ge 1$.

\par 

Equation \eqref{ME_reduced} is equivalent to a Fokker-Planck equation for the Wigner quasiprobability distribution function $\textbf{W}(\alpha)$~\cite{ghosh2017catalysis},
\begin{gather}
\frac{\partial \textbf{W}}{\partial t}=\left(\frac{\partial}{\partial\alpha}d_{\alpha}+\frac{\partial}{\partial\alpha^*}d_{\alpha}^{*}\right)\textbf{W}
+\left(D+\frac{\Gamma}{2}\right)\frac{\partial^{2}\textbf{W}}{\partial\alpha\partial\alpha^*},\label{eq:FP-sqz}
\end{gather}
where $d_{\alpha}=\frac{\Gamma}{2}\alpha-\kappa \alpha^{*}$. For an initial coherent state $|\alpha(0)\rangle$ of P, corresponding to the Wigner distribution $\textbf{W}(\alpha,\alpha^{*},0)=\frac{2}{\pi}e^{-2|\alpha-\alpha(0)|^{2}}$, the solution reads
\begin{gather}
\textbf{W}(x_1,x_2,t)={1\over 2\pi \sqrt{f_+f_-}}\exp\left\{-{1\over 2f_+f_-}\left[f_-(
x_1-x_{10}e^{\Gamma_+ t})^2 
+f_+(x_2-x_{20}e^{\Gamma_- t})^2\right]\right\},
\label{eq:Wig-sqz-coh}
\end{gather}
where $\Gamma_{\pm}=-\Gamma/2 \pm |\kappa|$. Here the real variables $x_{1}$, $x_{2}$ are defined by $\alpha =x_1+ix_2$, $x_{10} = \mathrm{Re}[\alpha(0)]$, $x_{20} = \mathrm{Im}[\alpha(0)]$, 
and the coefficients $f_\pm$ are the Gaussian widths
\begin{gather}
\quad\quad f_{\pm}=\frac{e^{2\Gamma_{\pm} t}}{4}+{\left(D+\frac{\Gamma}{2}\right)\over 4\Gamma_{\pm}}(e^{2\Gamma_{\pm} t}-1).
\label{eq:Wig-sqzcoh}
\end{gather}
Hence under quadratic pumping in the gain regime $\Gamma<0$, the initial coherent-state distribution evolves to a Gaussian with maximal and minimal widths $f_+$ and $f_-$ along the orthogonal axes $x_1$ and $x_2$ that are determined by the phase of the pump. The maximal width $f_+$ grows much faster than the minimal width $f_-$ (Fig.~\ref{fig:14}b), resulting in enhanced squeezing of the distribution.

\subsection*{Work extraction under non-linear pumping}

The maximum  extractable work from $\rho_\mathrm{P}$ is its ergotropy
\begin{equation}
\mathcal{W}(\rho_\mathrm{P})= \langle H_\mathrm{P}(\rho_\mathrm{P}) \rangle - \langle H_\mathrm{P}(\pi_\mathrm{P})\rangle.
\label{eq:maxw}
\end{equation}
Here $\pi_\mathrm{P}$ is the corresponding passive state of P. The passive state $\pi_\mathrm{P}$ corresponding to the Gaussian $\rho_\mathrm{P}$ is the thermal (Gibbs) state
\begin{equation}
\pi_\mathrm{P}(t)=Z^{-1}e^{-\frac{H_\mathrm{P}}{T_\mathrm{P}(t)}}
\label{eq:rpt}
\end{equation}
with slowly varying temperature $T_\mathrm{P}(t)$. We then find from Eqs.~\eqref{eq:maxw} and~\eqref{eq:rpt} the maximal power
\begin{gather}
\mathcal{P}_{\mathrm{Max}}=\dot{\mathcal{W}} -\dot{W}_{\mathrm{pump}}
=\dot {\langle H_{\mathrm{P}}\rangle} -T_\mathrm{P}(t)  \dot{\mathcal{S}}_\mathrm{P}(t)-\dot{W}_{\mathrm{pump}}.
\label{eq:pnonpas}
\end{gather}
Equation~\eqref{eq:pnonpas} is the net rate of extractable work converted from heat. The first term $\dot {\langle H_{P}\rangle}$ is the power obtained for a perfectly non-passive state, the second term $-T_\mathrm{P}(t) \mathcal{\dot{S}}_\mathrm{P}$ in~\eqref{eq:pnonpas} expresses its passivity increase due to the rise of the temperature $T_\mathrm{P}(t)$ and the entropy ${\mathcal{S}}_\mathrm{P}$ of P, and the third term is the subtracted power supplied by the pump. 

\par 

\begin{figure*}
	\centering
         \includegraphics[width=\columnwidth]{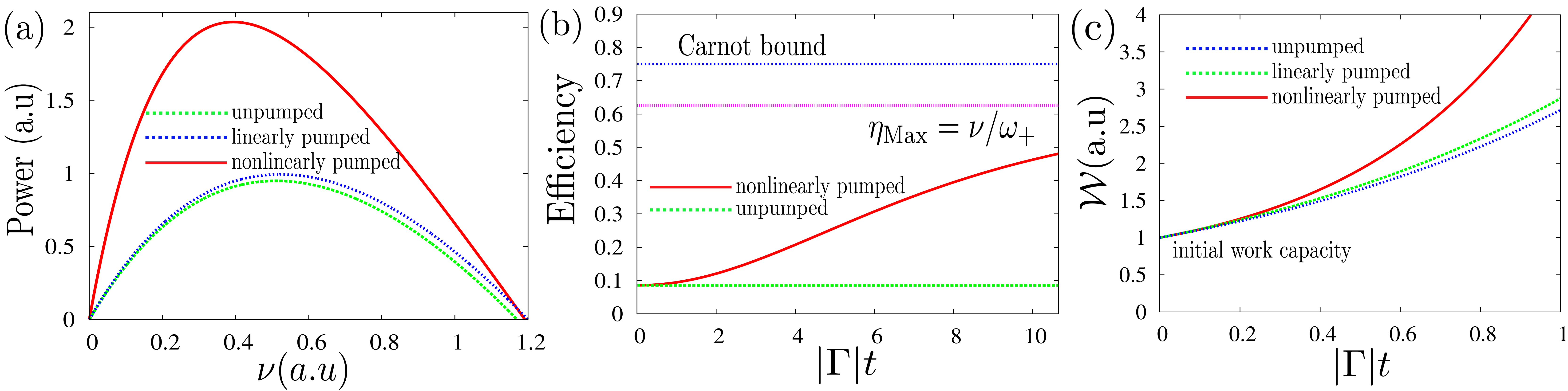}		 
\caption{(a)~Output power as a function of the piston frequency for quadratic pumping, linear pumping and without pumping, respectively for $T_{\mathrm{c}}=0.6T_{\mathrm{h}}$. (b)~Comparison of efficiency in the presence of weak non-linear pumping ($\kappa\neq 0,\;\text{and} \;|\kappa|/|\Gamma|\sim 0.1$) and in absence of any pumping ($\kappa=0$). (c)~The maximal extractable work (ergotropy) drastically increases in the presence of quadratic pumping compared to its linear and unpumped counterparts  (normalized by the initial work capacity) as a function of $|\Gamma|t$ for an initial coherent state with $|\alpha(0)|^{2} \sim 1$.}\label{fig:Fig2pnas-new}
\end{figure*}

As we show, the power may be strongly catalyzed by the pump squeezing (Fig.~\ref{fig:Fig2pnas-new}a). To obtain a better insight into this catalysis, we compute the engine efficiency bound which is defined as the ratio of the maximal net power output to the heat flux input $\dot{\mathcal{Q}}_{\mathrm{SP/h}}$,
\begin{gather}
\eta_{\mathrm{Max}}= \frac{\dot{\mathcal{W}} -\dot{W}_{\mathrm{pump}}}{\dot{\mathcal{Q}}_{\mathrm{SP/h}}}=\frac{\langle \dot{H}_{P}\rangle-T_\mathrm{P} \dot{\mathcal{S}}_\mathrm{P}-\dot{W}_{\mathrm{pump}}}{\dot{\mathcal{Q}}_{\mathrm{SP/h}}},
\label{eq:etaQHE}
\end{gather}
where the heat current from the hot bath to S+P is given by \cite{ghosh2017catalysis}
\bea
\dot{\mathcal{Q}}_{\mathrm{SP/h}} 
=-\omega_+ \Gamma\langle b^{\dag}b\rangle + \omega_+ D.\label{exprqsp}
\eea
Using Eq.~\eqref{ME_reduced} one finds
\bea
\dot{\langle H_\mathrm{P} \rangle } &=& \nu {\rm Tr} [\dot{\rho}_\mathrm{P} b^{\dag} b ] 
 = -\nu \Gamma\langle b^{\dag} b\rangle +\nu D 
+\nu \kappa \langle b^{\dag 2}\rangle +\mathrm{c.c.},\label{exprhp}\\
\dot{W}_{\mathrm{pump}}(t) &=& {\rm Tr}[\rho_\mathrm{P} \frac{d}{dt}H_{\mathrm{pump}}(t)] 
= \nu \kappa \langle b^{\dag 2}\rangle + \mathrm{c.c.}\label{exprwp}
 \eea
Combining Eqs.~\eqref{exprqsp}-\eqref{exprwp}, we finally arrive at
\be
\dot{\langle H_\mathrm{P} \rangle} -\dot{W}_{\mathrm{pump}} = \frac{\nu}{\omega_+} \dot{\mathcal{Q}}_{\mathrm{SP/h}}\label{Hp-Wpump}.
\ee

\subsection*{Efficiency boost}
From Eqs.~\eqref{eq:etaQHE} and~\eqref{Hp-Wpump}, the efficiency bound can be expressed as
\bea
\eta_{\mathrm{Max}} &=& \frac{\nu}{\omega_{+}} - \frac{\nu\dot{n}_{\mathrm{pas}}}{\dot{\mathcal{Q}}_{\mathrm{SP/h}}},
\label{supercarnot}
\eea
where we have used the identity $T_\mathrm{P}\dot{\cal S}_\mathrm{P}=\nu\dot{n}_{\mathrm{pas}}$ for Gaussian states that 
relates the entropy increase to $n_{\mathrm{pas}} = (e^{\nu/T_\mathrm{P}} - 1)^{-1}$, the thermal excitation of the passive state~\eqref{eq:rpt}. Equation~\eqref{supercarnot} shows that the efficiency depends on the \textit{ratio of $\nu\dot{n}_{\mathrm{pas}}$ to the incoming heat flow}. In turn, $\nu\dot{n}_{\mathrm{pas}}$ and $\dot{\mathcal{Q}}_{\mathrm{SP/h}}$ depend on the evolving $T_{\mathrm{P}}(t)$ and squeezing parameter $r(t)$~\cite{ghosh2017catalysis} of P, and on the expectation values $x_{10}$, $x_{20}$, of the quadrature operators $\hat{x}_1$ and $\hat{x}_2$ in the initial state of P.

\par

We can simplify the exact expression of $\eta_{\mathrm{Max}}$ using the parametrization of Gaussian states in terms of the squeezing parameter $r(t)$~\cite{ghosh2017catalysis} which relates to the quadrature widths $f_{\pm}$~[cf.~\eqref{eq:Wig-sqzcoh}] of the distribution~\eqref{eq:Wig-sqz-coh},
\bea
(n_{\mathrm{pas}} + 1/2)\cosh{2r(t)} = f_++f_-.
\eea
Then one obtains~\cite{ghosh2017catalysis}
\bea
\dot{n}_{\mathrm{pas}} &=& -\Gamma(n_{\mathrm{pas}}+1/2) + (D+\Gamma/2)\cosh{2r(t)},\nn\\
\dot{\mathcal{Q}}_{\mathrm{SP/h}} 
 &=& \omega_+(D+\Gamma/2) -\omega_+\Gamma\left[(n_{\mathrm{pas}}+1/2)\cosh{2r(t)}+ x_{10}^2e^{2\Gamma_{+}t} + x_{20}^2e^{2\Gamma_{-}t}\right].\label{nthQ}
\eea 
Both $\dot{n}_{\mathrm{pas}}$ and the heat flow are enhanced by the squeezing, but remarkably, the heat flow $\dot{\mathcal{Q}}_{\mathrm{SP/h}}$ is more strongly enhanced, which yields an ergotropy increase along with an efficiency increase.

\par 

Assuming that $n_{\mathrm{pas}}\gg D/|\Gamma|$, for any Gaussian (squeezed) state the efficiency can be computed as~\cite{ghosh2017catalysis}
\be
\eta(t) \simeq  \frac{\nu}{\omega_{+}}\left[1- \frac{n_{\mathrm{pas}}+1/2}{(n_{\mathrm{pas}}+\frac{1}{2})\cosh{2r(t)} + (x_{10}^{2}e^{2\Gamma_+ t}+ x_{20}^2e^{2\Gamma_- t}) }\right]\label{eq:eta_non}.
\ee 
The efficiency reaches the maximal attainable efficiency $\eta_{\mathrm{Max}}$, bounded by the Carnot efficiency~\cite{ghosh2017catalysis},
\begin{gather}
\eta_{\mathrm{Max}} 
:=\frac{\nu}{\omega_+} \leq \eta_{\mathrm{Carnot}}= 1-\frac{T_{\mathrm{c}}}{T_{\mathrm{h}}}.
\label{eq:carnotmaser}
\end{gather}
It is seen from Fig.~\ref{fig:Fig2pnas-new}b and the analysis of the expression~\eqref{eq:eta_non} that $\eta$ tends to $\eta_{\mathrm{Max}}=\frac{\nu}{\omega_{+}}$ as the pumping rate $\kappa$ increases. Thus $\eta_{\mathrm{Max}}$ can approach the Carnot efficiency by choosing $T_\mathrm{h}/T_\mathrm{c}=\omega_+/\omega_0$. 

\par 

When the pumping is off ($\kappa =r(t) =0$), the passivity term that limits the ergotropy~\eqref{eq:maxw} or the power~\eqref{eq:pnonpas} becomes small only in the semiclassical limit $x_{10}^{2}+x_{20}^{2}=|\alpha(0)|^2 \gg 1$  and under the weak coupling condition $(g/\nu)|\alpha(0)| \ll 1$. The efficiency in the unpumped gain regime $\Gamma<0$ is then
\bea
\eta_{0} 
= \frac{\nu}{\omega_{+}}\left[\frac{|\alpha(0)|^2}{|\alpha(0)|^2-D/\Gamma}\right]=\frac{\nu}{\omega_{+}}\frac{1}{1+\frac{D}{|\Gamma||\alpha(0)|^2}}.
\label{eta0_exact}
\eea

\par

A comparison between the unpumped case~\eqref{eta0_exact} and the pumped case~\eqref{eq:eta_non} and~\eqref{eq:carnotmaser} shows that quadratic pumping may dramatically enhance the efficiency (Fig.~\ref{fig:Fig2pnas-new}b) and the ergotropy as shown in Fig.~\ref{fig:Fig2pnas-new}c for an initial piston charging $|\alpha(0)|^{2} \sim 1$. The reason is that when the non-linear pumping is on, any heat input in $P$ is amplified by the squeezing as $\nu(n_{\mathrm{pas}}+1/2)\cosh{2r(t)}$ which enhances the leading term in the denominator of~\eqref{eq:eta_non}. On the other hand, the terms depending on $\Gamma_\pm$ therein [cf.~\eqref{eq:Wig-sqz-coh}] and the passive energy, $\nu n_{\mathrm{pas}}$, are unaffected by the squeezing. As a consequence, the stronger the squeezing, the higher the efficiency. 

\par 

Similar calculations lead to the following work capacity for its \textit{linear pumping} counterpart (pump Hamiltonian $H_{\mathrm{pump}}(t) = i\kappa b^{\dag} e^{-i\nu t} + \mathrm{h.c.}$),
\begin{gather}
\mathcal{W}_{\mathrm{L}}=\nu|\alpha(t)|^{2},
\label{eq:wlin}
\end{gather}
where $\alpha(t)$ is the generated phase-space displacement $\alpha(t):=\alpha(0)e^{-\frac{\Gamma t}{2}}+\frac{2\kappa}{|\Gamma|}(e^{-\frac{\Gamma t}{2}}-1)$. Linear pumping generates an energy contribution which is additive to the passive energy, $\nu n_{\mathrm{pas}} + \nu |\alpha(t)|$. The efficiency for linear pumping is found to be limited by 
\begin{gather}
\eta_\mathrm{L} \xrightarrow[t \gg |\Gamma|^{-1}]{} \frac{\nu}{\omega_+} \frac{|\alpha(0) +2\frac{\kappa}{|\Gamma|}|^2}{|\alpha(0) +2\frac{\kappa}{|\Gamma|}|^2 + n_{\mathrm{pas}}(0) +D/|\Gamma|}.
\end{gather}
This efficiency never approaches $\eta_{\mathrm{Max}}=\frac{\nu}{\omega_+}$ if $|\alpha(0)|$ is much smaller than $n_{\mathrm{pas}}(0) +D/|\Gamma|$. Consequently, the ergotropy increase generated by heat input for linear pumping is much less significant than for non-linear pumping resulting in much lower power.

\section{Discussion: quantumness and engine performance}\label{sec_quantum}

We have shown that the operation of machines powered by quantum-coherent (non-thermal) baths or catalyzed by quantum-non-linear pumping of the piston crucially depends on whether the working fluid (WF) or the piston are in a passive or non-passive state. However, non-passivity also exists in a classical context~\cite{gorecki1980passive,daniels1981passivity,daprovidencia1987variational} which prompts the question: To what extent is the performance of these machines truly affected by quantum features of the bath, the WF and (or) the piston? This question requires the clarification of two points: (i) What are the relevant criteria for the bath, WF or piston quantumness? (ii) Is there a compelling link between such quantumness and the machine performance? (iii) Conversely, does the fact that a machine with quantum ingredients is fueled by a heat bath imply that the machine conforms to the rules of thermal (heat) engines?

\par 

\begin{figure}
	\centering
         \includegraphics[width=\columnwidth]{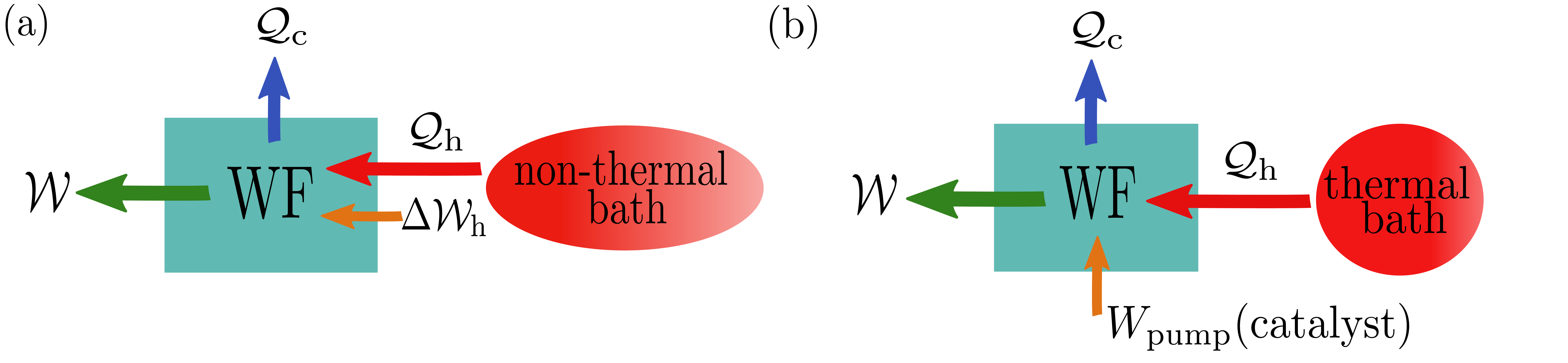}	 
\caption{(a) Scheme of a machine powered by a squeezed (non-thermal) bath \cite{rossnagel2014nanoscale,niedenzu2016operation,dag2016multiatom}. (b) Scheme of a catalyzed quantum heat machine.}
	\label{fig:final}
\end{figure}

\par

As long as the dynamics of the harmonic-oscillator (HO) WF or piston is described by linear or quadratic operators, the Gaussian character of their  state is preserved~\cite{wallquist2010single}. A Gaussian HO state is non-classical if its $P$-function is negative~\cite{glauber1963coherent,dodonov2002nonclassical,gardinerbook}. According to this criterion, a squeezed thermal distribution with thermal photon number $\nbar$ and squeezing parameter $r>0$ is non-classical only if its fluctuations are below the minimum uncertainty limit. This holds if $\nbar<(e^{2r}-1)/2$~\cite{kim1989properties,gardinerbook}.

\par

Whether or not the WF or the piston are in a non-classical state, however, has no direct impact on the machine's operation ---\emph{only the energy and ergotropy of their state play a role}. The highest (near-unity) efficiency is attained by nearly \textit{mechanical operation} of the machine. It is effected by the ergotropy imparted by a squeezed bath to the WF or by a non-linear pump to the piston (Fig.~\ref{fig:final}). Thus the possibility of strong catalysis of heat-to work conversion arises from ergotropy enhancement as the piston undergoes squeezing after it is initialized in any nearly-coherent state and subjected to quadratic pumping. Yet, non-classicality and non-passivity do not generally go hand-in-hand: Coherent thermal states are classical but non-passive, whereas squeezed thermal states may be either classical or non-classical.

\par 

In general for Gaussian and non-Gaussian states alike, the machine performance is optimized for a WF or piston state with the highest possible ergotropy allowed for its energy. Equivalently, the quantumness of an engine and bath may be deemed useful, if given a certain energy to transfer (eg. squeeze) a thermal bath, the unitary transformation that maximizes the ergotropy of the bath has no classical counterpart. Thus, neither the \emph{operational principles} of a cyclic machine nor its performance demand the non-classicality of the state of its ingredients. The extracted work and efficiency are optimized by maximizing the ergotropy and minimizing the passive (thermal) energy of the WF and/or the piston, but are not necessarily related to the non-classicality of their state.

\bigskip

ACKNOWLEDGEMENTS: We acknowledge discussions with A. G. Kofman and support from ISF, BSF and VATAT. W.\,N.\ acknowledges support from an ESQ fellowship of the Austrian 
Academy of Sciences (\"OAW).

\end{document}